 \documentclass[12pt,preprint]{aastex}

\usepackage{epsf}

\newcommand{\mjb}{mJy~beam$^{-1}$}
\newcommand\jb{Jy~beam$^{-1}$}
\newcommand\kms{km~s$^{-1}$}

\shorttitle{OH (1720 MHz) Masers in SNR W28}
\shortauthors{Hoffman et al.}

\begin{document}

\title{The OH (1720 MHz) Supernova Remnant Masers in W28: \\ MERLIN and VLBA Polarization Observations}

\author{Ian M.\ Hoffman}
\affil{National Radio Astronomy Observatory, P.\ O.\ Box O, Socorro, NM, USA 87801}
\affil{Department of Physics and Astronomy, University of New Mexico, Albuquerque, NM, USA 87131}
\email{ihoffman@nrao.edu}

\author{W.\ M.\ Goss}
\affil{National Radio Astronomy Observatory, P.\ O.\ Box O, Socorro, NM, USA 87801}

\author{C.\ L.\ Brogan}
\affil{Institute for Astronomy, 640 North A`ohoku Place, Hilo, HI, USA 96720}

\and

\author{M.\ J Claussen}
\affil{National Radio Astronomy Observatory, P.\ O.\ Box O, Socorro, NM, USA 87801}

\begin{abstract}
Full-polarization MERLIN and VLBA observations of the 1720-MHz maser emission from the OH molecule in the supernova remnant W28 are presented.
Zeeman splitting ($|\mathbf{B}| \approx 0.75$~mG) has been directly resolved between right- and left-circularly polarized spectra for the first time.
Linear-polarization position angles and circular-polarization Zeeman splitting observed in the maser emission permit interpretation of a comprehensive picture of the magnetic field at the supernova/molecular cloud interface marked by the masers.
We find the post-SNR-shock magnetic field to be well-ordered over the $\sim 1$~pc covered by the maser region (compared with the approximately 30-pc diameter of the entire SNR) and well-aligned with the shock front that is traced by synchrotron radiation and molecular emission.
Based on MERLIN data having a resolution of 200~mas and VLBA data with a resolution of 15~mas, the masers are measured to have deconvolved angular sizes of 90 to 350~mas (225 to 875~AU) with compact cores 20~mas (50~AU) in size, consistent with theoretical expectation and previous observations.
\end{abstract}


\section{Introduction}

Spectral-line emission from the OH (1720 MHz) satellite transition $({^2}{\Pi}_{3/2},J={3\over2},F=2{\rightarrow}1)$ in supernova remnants (SNR's) was first observed by Goss (1968) toward W28.
The maser nature of the emission was noted shortly thereafter along with the recognition that the 1720-MHz line emission, when observed in association with absorption in the other ground-state OH lines, requires a pump mechanism different than that used to explain \ion{H}{2} region OH masers (Goss \& Robinson 1968; Ball \& Staelin 1968; Turner 1969; Robinson, Goss, \& Manchester 1970).
Much later, observations of W28 by Frail, Goss, and Slysh (1994) renewed the interest in the SNR masers while also recognizing (1) the astrophysical diagnostic value of the masers' association with molecular clouds and (2) the viability of the collisional pump suggested by Elitzur (1976) to explain the level population inversion.
Subsequent studies and surveys (Yusef-Zadeh et al.\ 2000; Yusef-Zadeh et al.\ 1999; Koralesky et al.\ 1998; Green et al.\ 1997; Frail et al.\ 1996; Yusef-Zadeh et al.\ 1996) have since observed approximately 200 Galactic supernova remnants and found 22 remnants with associated OH (1720~MHz) maser emission.

\begin{figure}
\hbox{ \epsfxsize=4.50in \epsfbox{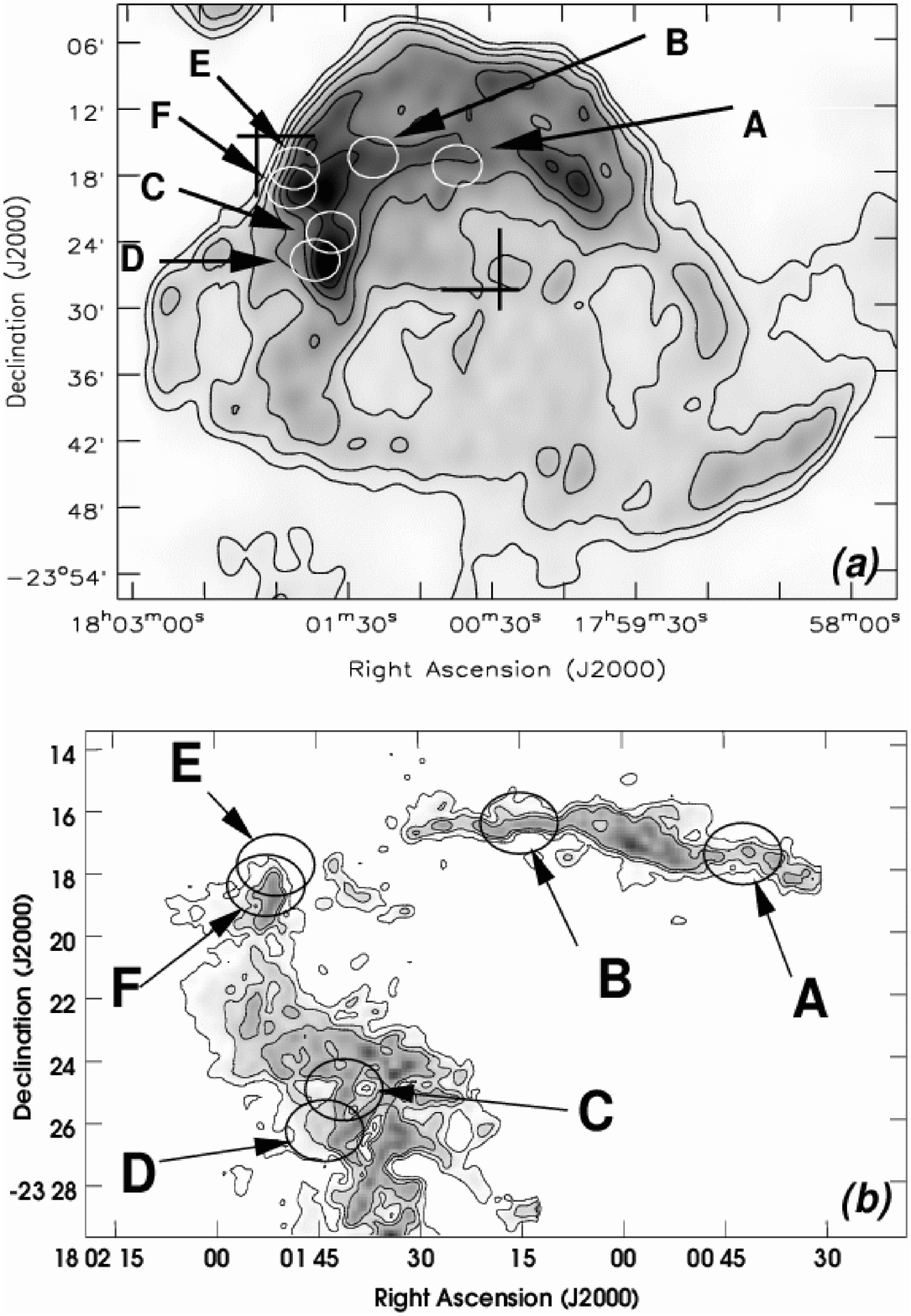}}
\caption{{\it (a)} A contour and greyscale VLA image of the 330 MHz continuum radiation from W28 (C.\ Brogan 2004, private communication).
The contour levels are 0.2, 0.5, 1, 1.5, 2, 4, and 6~\jb\ and the beam is $2\farcm7 \times 1\farcm7$ at a position angle of $-11$\arcdeg.
The crossed lines mark the southwest and northeast corners of panel {\it (b)}.
{\it (b)} A contour and greyscale, moment-zero image of the CO ($J = 3 \rightarrow 2$, 346~GHz) emission from W28 observed by Arikawa et al.\ (1999) using the James Clerk Maxwell Telescope, indicating the post-shock gas.
The contours are 10, 20, and 30 times the image RMS noise level of $2.3 \times 10^4\ {\rm K}\,{\rm m}\,{\rm s}^{-1}$.
The beam is 15\arcsec.
The OH (1720~MHz) maser regions are identified following the nomenclature of C97.
The sizes of the ellipses used to to indicate the maser regions have no special significance; the regions contain between three and 17 masers as observed using the VLA (C97).
\label{c97+CO}}
\end{figure}

Observations support the premise that OH (1720~MHz) SNR masers occur where the shock front from a supernova explosion encounters a molecular cloud ({\it e.g}.\ W28:  Frail, Goss, and Slysh 1994; Kes~78:  Koralesky et al.\ 1998; 3C391:  Frail et al.\ 1996; W44: Wootten 1977).
In such interactions, a $C$-type (non-dissociative) shock can produce the relatively rare conditions ($n \approx 10^5\,{\rm cm}^{-3}$, $T \approx 90$~K, $N_{\rm OH} \sim 10^{16-17}\,{\rm cm}^{-2}$) needed for the collisionally-pumped OH (1720 MHz) masers (Elitzur 1976; Wardle 1999; Lockett, Gauthier, \& Elitzur 1999; see also Draine \& McKee 1993).
The suggested temperature, density, and OH column of the OH (1720~MHz) collisional pump are consistent with those determined from observation ({\it e.g}.\ Claussen et al.\ 1997, hereafter C97; Hoffman et al.\ 2003a, hereafter H03).  
Furthermore, OH (1720~MHz) SNR masers are observed at the SNR/molecular cloud interface where the shock front is edge-on, moving transversely across the plane of the sky ({\it e.g}.\ Reynoso \& Mangum 2000).
Frail \& Mitchell (1998) and Arikawa et al.\ (1999) have examined the post-shock gas in W28 using observations of shocked CO ($J = 3 \rightarrow 2$) using the James Clerk Maxwell Telescope.
Figure \ref{c97+CO} shows the good positional agreement between the OH masers, the shocked gas, and the limb of the synchrotron emission from the SNR (this agreement is discussed in \S\ref{lin_disc}).

The intrinsic, transverse sizes of the masers have been the subject of several recent studies.
Indeed, the narrow-band maser emission does not permit simple multi-wavelength fitting of the scattering properties to the $\lambda^2$ wavelength dependence expected of scattering due to the ionized interstellar medium at radio frequencies.
Claussen et al.\ (2002) suggest that the angular scatter-broadening of the maser images in W28 is not significant based on studies of the angular scattering of an adjacent extragalactic source and the temporal broadening of a background pulsar.
Hoffman et al.\ (H03) observed the OH (1720~MHz) SNR masers in IC~443 near the Galactic anticenter where interstellar scatter-broadening is expected to be negligible.
Conversely, Yusef-Zadeh et al.\ (1999) find the OH (1720~MHz) SNR masers in the direction of the Galactic center to be appreciably scatter-broadened.
Both Claussen et al.\ (2002) and H03 find that the observed sizes of the masers are consistent with those predicted from collisional pump theory (${\sim}10^{15}$~cm, Lockett et al.\ 1999).
However, OH (1720~MHz) SNR maser observations using the VLBA and MERLIN (H03; Brogan et al.\ 2003; Claussen et al.\ 1999, hereafter C99) are not sensitive to the very extended (60\arcsec), ``face-on'' maser emission observed in W28 using the VLA (Yusef-Zadeh, Wardle, \& Roberts 2003, $60\arcsec \approx 10^{18}$~cm projected linear size for a distance $d=2.5$~kpc to the SNR, Vel\'{a}zquez et al.\ 2002).
The current study will examine only the approximately 40 individual maser ``spots'' observed by C97 (using the VLA) at the location where the shock front of W28 is viewed edge-on (moving transversely across the sky).
The maser region observed for this paper is about 1~pc in diameter compared with the approximately 30-pc diameter of the W28 SNR (Fig.~\ref{c97+CO}).

The intrinsic angular size of OH (1720~MHz) SNR maser spots (on the order of 100~mas) is a relatively problematic angular scale for observation with most existing instruments.
Radio interferometers with baselines on the order of 10~km in length do not resolve the maser emission and, furthermore, typically convolve several maser spots together within a large beam (on the order of 1\arcsec).
This ``spatial blending'' confuses both the angular and spectral character of the individual masers, preventing a determination of intrinsic emission properties.
Past high-angular-resolution studies have employed VLBI-baseline lengths on the order of 1000~km (fringe spacings of approximately 30~mas).
Although OH (1720~MHz) SNR masers are known to have a compact core component (C99; H03), VLBI experiments typically ``resolve out'' most maser spots ({\it i.e}.\ the observations are not sensitive to emission on 100-mas angular scales due to low correlated fringe visibility).
However, the angular sensitivity of the MERLIN array (baselines on the order of 100~km in length) is well-matched to the angular scale of the masers.
The minimum angular resolution of observations using MERLIN (approximately 100~mas) is sufficient to alleviate spatial blending.
The largest angular scale sampled in observations using MERLIN (approximately 3\arcsec) is sufficient to recover all of the flux density of the masers.

Since the OH radical is paramagnetic, the polarization state of OH emission is strongly affected by the magnetic field in the maser gas (compared with non-paramagnetic molecules such as H$_2$CO).
Indeed, circular-polarization studies of the Zeeman splitting of OH spectral lines is a significant source of information about the interstellar magnetic field ({\it e.g}.\ Caswell 2004).
Zeeman observations of the OH (1720~MHz) maser emission in supernova remnants have provided estimates of the magnitude of the magnetic fields in supernova/molecular cloud interactions ($|\mathbf{B}| \approx 0.5$~mG: C99; Brogan et al.\ 2000, hereafter BFGT).
Also, observations of the linear polarization of the maser radiation can yield the orientation of the post-SNR-shock magnetic field.
However, the suggested theoretical conversion from observational parameters to the three-dimensional magnetic field geometry responsible for the observed maser polarization is not straightforward ({\it e.g}.\ Elitzur 1998; Watson \& Wyld 2001, hereafter WW01; Gray 2003).
Nevertheless, measurement of the following three observational polarization parameters is a necessary step toward understanding the physics of the maser emission:
(1) the magnitude of the level splitting due to the Zeeman effect,
(2) the position angle $\chi$ of the linear polarization,
and (3) the fractional linear polarization $q \equiv \sqrt{Q^2+U^2}/I$ of the maser radiation (where $Q$, $U$, and $I$ are Stokes parameters).

To date, all of the full-polarization studies of OH (1720~MHz) SNR masers were completed by C97 and BFGT using the VLA.
Since much of the prevailing theory pertinent to these masers was published after the C97 paper (Elitzur 1998; Lockett, Gauthier, \& Elitzur 1999; Wardle 1999), BFGT contains a thorough discussion of the theories and their implications for both the C97 and BFGT data.
The interpretation of many of the results of the current paper also rely on BFGT.
However, due to the insufficient spectral and angular resolution of the VLA observations presented in C97 and BFGT, the theoretical suggestions presented by BFGT could not be confidently constrained using the data available at the time.
Indeed, the observational hurdle currently confronting OH (1720~MHz) SNR maser studies is the acquisition of full linear- and circular-polarimetry of resolved maser sources.
The purpose of the current study is to observe a large sample of OH (1720~MHz) SNR masers with spectral and angular resolution sufficient to apply more definitively the wealth of maser polarization theory discussed by BFGT.
The W28 SNR is chosen for this purpose because of the large number of masers (approximately 40).

This paper presents full polarization observations of the OH (1720~MHz) masers in W28 made using the Multi-element Radio Linked Interferometry Network\footnote{MERLIN is operated as a National Facility by the University of Manchester, Jodrell Bank Observatory, on behalf of Particle Physics and Astronomy Research Council (PPARC).} (MERLIN) and the Very Long Baseline Array (VLBA) of the NRAO\footnote{The National Radio Astronomy Observatory (NRAO) is a facility of the National Science Foundation operated under a cooperative agreement by Associated Universities, Inc.}.
This paper will use the source numbering convention of Claussen et al.\ (C97 Table~1).

\section{Observations and Data Reduction}

\subsection{MERLIN+Lovell}

The W28 masers were observed on 29 and 31 January 2002 using the MERLIN radio telescope of Jodrell Bank Observatory for a total of approximately 8 hours.
The observations were centered on a Doppler velocity of $v_{\rm LSR} = 11$~\kms\ for a line rest frequency of 1720.52998~MHz.
Seven antennas were used; the Mark II and Lovell telescopes at Jodrell Bank, the 32~m antenna at Cambridge, and the 25~m dishes at Knockin, Darnhall, Tabley, and Defford.
The very short ($< 500$~m) Mark II-Lovell baseline is not used.
The baseline lengths of MERLIN range from 11~km to 217~km; the array is not sensitive to angular scales larger than 3\farcs3.
The antennas have right- (R) and left- (L) circularly polarized feeds from which RR, LL, RL, \& LR cross-correlations were formed.
The correlator produced 256 spectral channels across 250~kHz (44~\kms).
The spectra were Hanning-smoothed off-line yielding a velocity resolution of 0.36~\kms.
The visibilities were integrated for 8.0~s.

The absolute amplitude calibration is based on observations of 3C286.
The bandpasses were calibrated based on observations of 3C84.
The phases were calibrated using frequent observations at 14~MHz bandwidth of the source 1748-253 (J1751-2524), which has a flux density of 1.3~Jy and a 1-$\sigma$ positional uncertainty of 2.7~mas (Fomalont et al.\ 2003).
The visibility phases and amplitudes were then (self-) calibrated using images of the bright (approximately 70~Jy) F39 maser (C97).
3C84 was observed at both 14~MHz and 256~kHz bandwidth allowing bandwidth-dependent calibration to be transferred.
The position angle of the linear polarization response of the antennas was determined from the observations of 3C286 and 1748-253.
The RMS noise in the final single-channel images is 8~\mjb, in agreement with expected instrumental behavior.
The FWHM synthesized beam of the images is $550\times100$~mas at a position angle $9{\arcdeg}$.

The tracking position during the observations was the position of the F39 maser determined from VLBA observations by C99 ($\alpha_{\rm J2000}= 18^{\rm h}01^{\rm m}52\fs7054$, $\delta_{\rm J2000}= -23\arcdeg19\arcmin24\farcs641$).
Since the primary beam of the largest MERLIN antenna (the 76~m Lovell) is approximately 10\arcmin, only the E and F region masers are within the primary beam of the array and the amplitude of the A, B, C, and D region masers are attenuated (the A region lies in a sidelobe of the primary beam of the Lovell antenna).
Also, the relatively large angular distance from the phase tracking center to the outlying maser regions resulted in decreased amplitude due to decoherent fringe visibility averaging during on-line integration (`time smearing'; see Thompson 1999).
These amplitude degradations have been corrected for the few masers detected in the outlying regions.

\subsection{VLBA+Y1}

We also observed the W28 OH (1720~MHz) masers on 26 September, 7 and 14 October 2002 and 24 February, 14 and 24 March 2003 with the ten antennas of the VLBA plus one antenna of the VLA (Y1) for approximately 30~hours.
The antennas have right- (R) and left- (L) circularly polarized feeds from which RR, LL, RL, \& LR cross-correlations were formed.
The correlator produced 128 spectral channels across a 62.5~kHz (11~\kms) band yielding a 0.09~\kms\ velocity channel spacing.
The spectra were off-line Hanning-smoothed yielding a velocity resolution of 0.18~\kms.
Two overlapping observing bands were used, centered on $v_{\rm LSR} = 13$~\kms\ and $v_{\rm LSR} = 8$~\kms\ in order to cover the 16~\kms\ range over which the maser are observed.
The visibilities were integrated for 4.7~s.

The amplitude scale is set using online system temperature monitoring and {\it a priori} antenna gain measurements.
Bandpass responses and station delays were found from observations of 3C345.
The visibility phases and amplitudes were then (self-) calibrated using images of the bright (approximately 70~Jy) F39 maser (C97).
The position angle of the linear polarization response of the antennas is determined from observations of 3C286 and J1751+0939.

The baseline lengths of the VLBA+Y1 array range from 52~km to 8611~km; the array is not sensitive to angular scales larger than 0\farcs70.
The correlated field of view with respect to the phase-tracking position is $30\arcsec$.
Thus, in order to observe all of the W28 masers, two array-pointing positions and eight correlator-tracking positions were used, listed in Table~1.
The synthesized beam of the resulting images is $25\times 9$~mas at a position angle of $10{\arcdeg}$.
The RMS noise in the final single-channel images is 30~\mjb, in agreement with expected values.

\section{Results}

\subsection{Angular Structure and Position}

\begin{figure}
\hbox{ \epsfxsize=2.50in \epsfbox{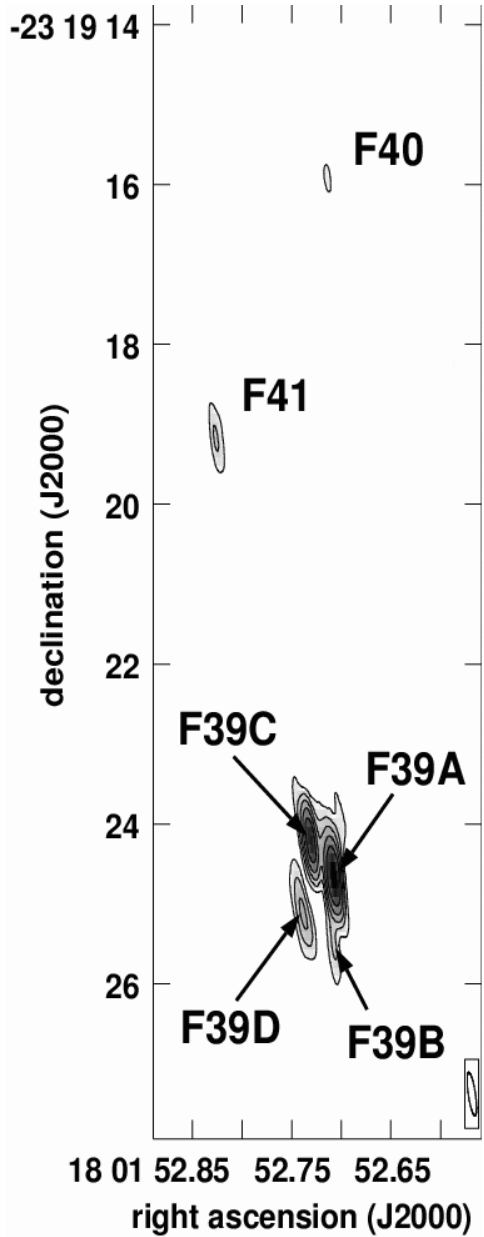}}
\caption{A line-integrated (moment-zero) contour and greyscale image of part of the W28 F maser region from the MERLIN observations showing the resolved angular structure of the F39 maser.
The contours are 10, 12, 25, 50, 100, 200, and 500 times $222\,{\rm Hz}\,{\rm Jy}\,{\rm beam}^{-1}$.
The beam, plotted in the lower right-hand corner, is $550\times100$~mas at a position angle of $9{\arcdeg}$.
\label{moment0}}
\end{figure}

\begin{figure}
\hbox{ \epsfxsize=4.50in \epsfbox{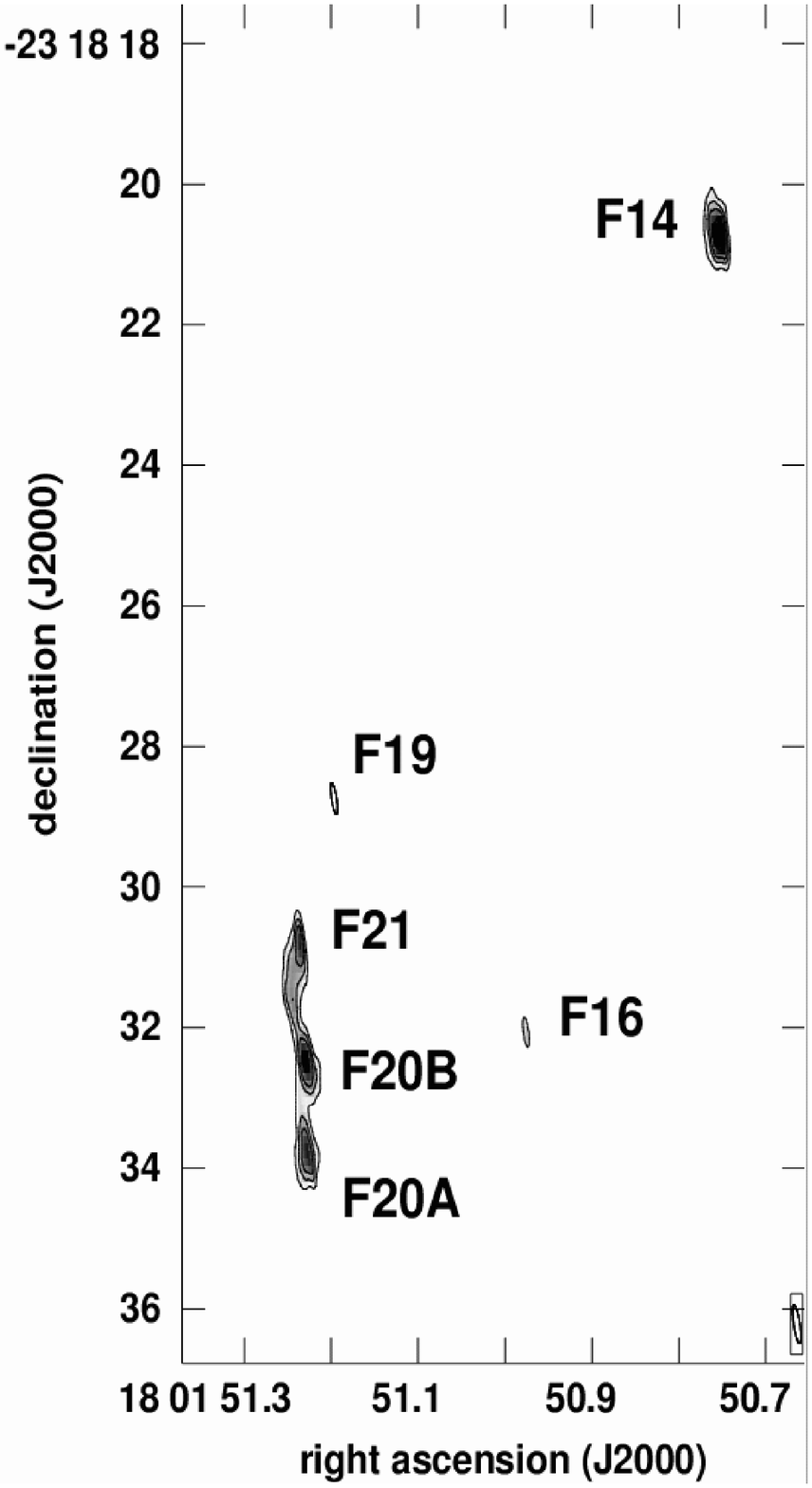}}
\caption{A line-integrated (moment-zero) contour and greyscale image of part of the W28 F maser region from the MERLIN observations.
The contours are 10, 15, 20, and 30 times $180\,{\rm Hz}\,{\rm Jy}\,{\rm beam}^{-1}$.
The beam, plotted in the lower right-hand corner, is $550\times100$~mas at a position angle of $9{\arcdeg}$.
\label{moment0-3}}
\end{figure}

\begin{figure}
\hbox{ \epsfxsize=4.00in \epsfbox{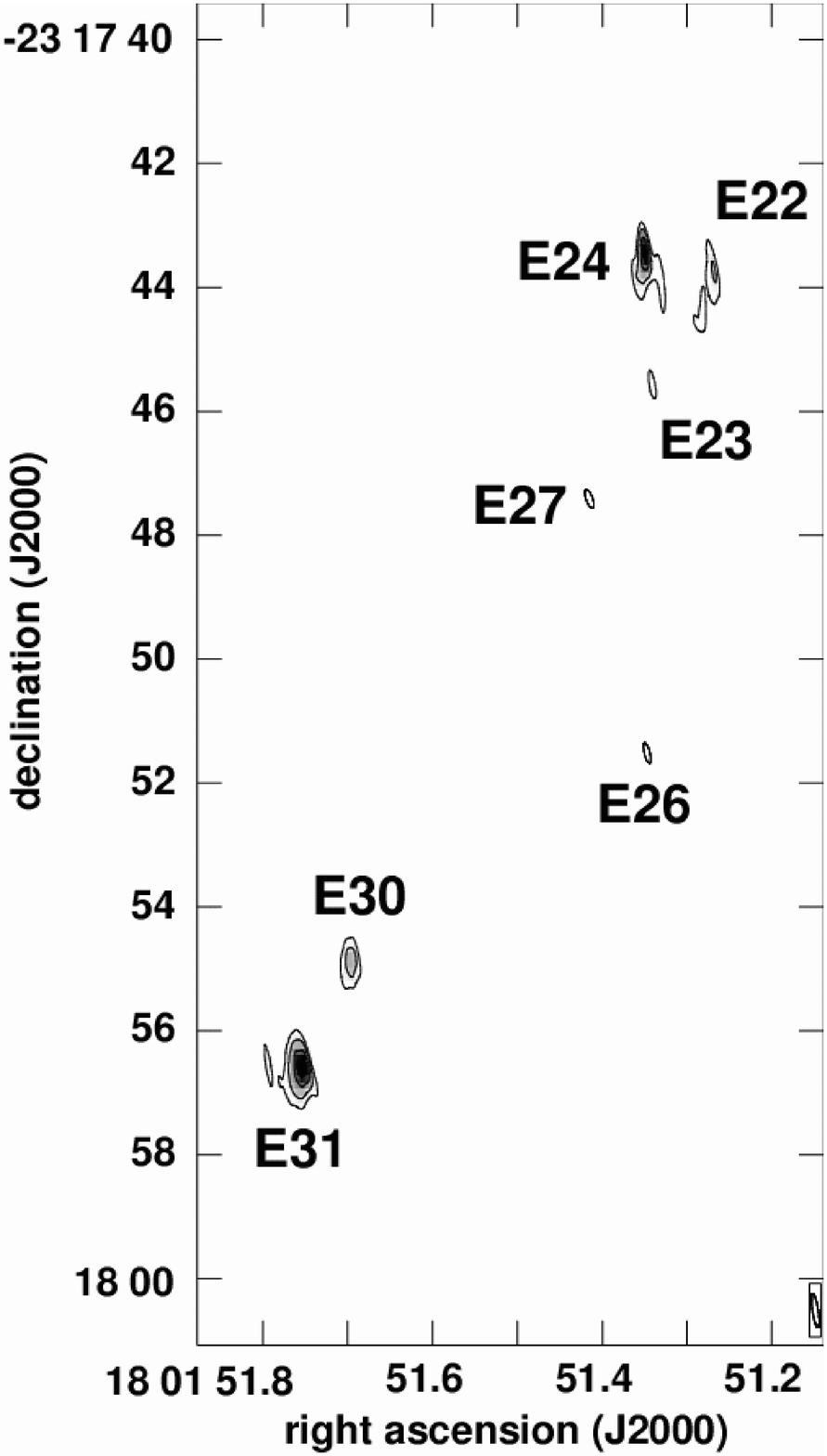}}
\caption{A line-integrated (moment-zero) contour and greyscale image of part of the W28 E maser region from the MERLIN observations.
The contours are 15, 30, 60, and 90 times $180\,{\rm Hz}\,{\rm Jy}\,{\rm beam}^{-1}$.
The beam, plotted in the lower right-hand corner, is $550\times100$~mas at a position angle of $9{\arcdeg}$.
\label{moment0-1}} 
\end{figure}

\begin{figure}
\plotone{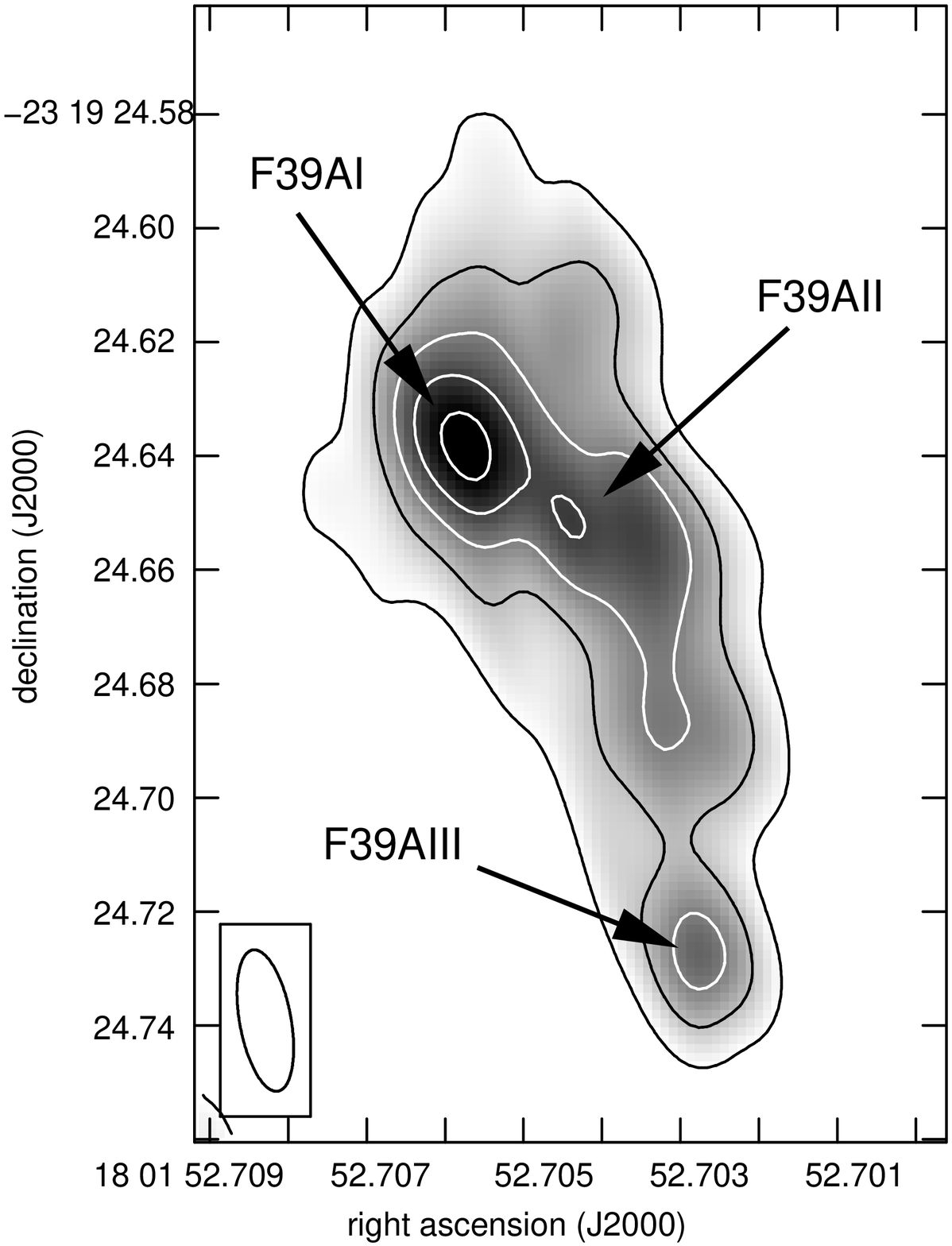}
\caption{Contour and greyscale image of the F39A maser region using the $v_{\rm LSR} = 11.2$~\kms\ channel from the VLBA observations.
The contour levels are -15, 15, 30, 45, 60, and 75 times image RMS noise level of 30~\mjb\ (no negative contours appear).
The beam, plotted in the lower left-hand corner, is $25 \times 9$~mas at a position angle of 10\arcdeg.
The three image features are labelled (Table~\ref{vlbapos}).
\label{VLBA_F39A}}
\end{figure}

\begin{figure}
\plotone{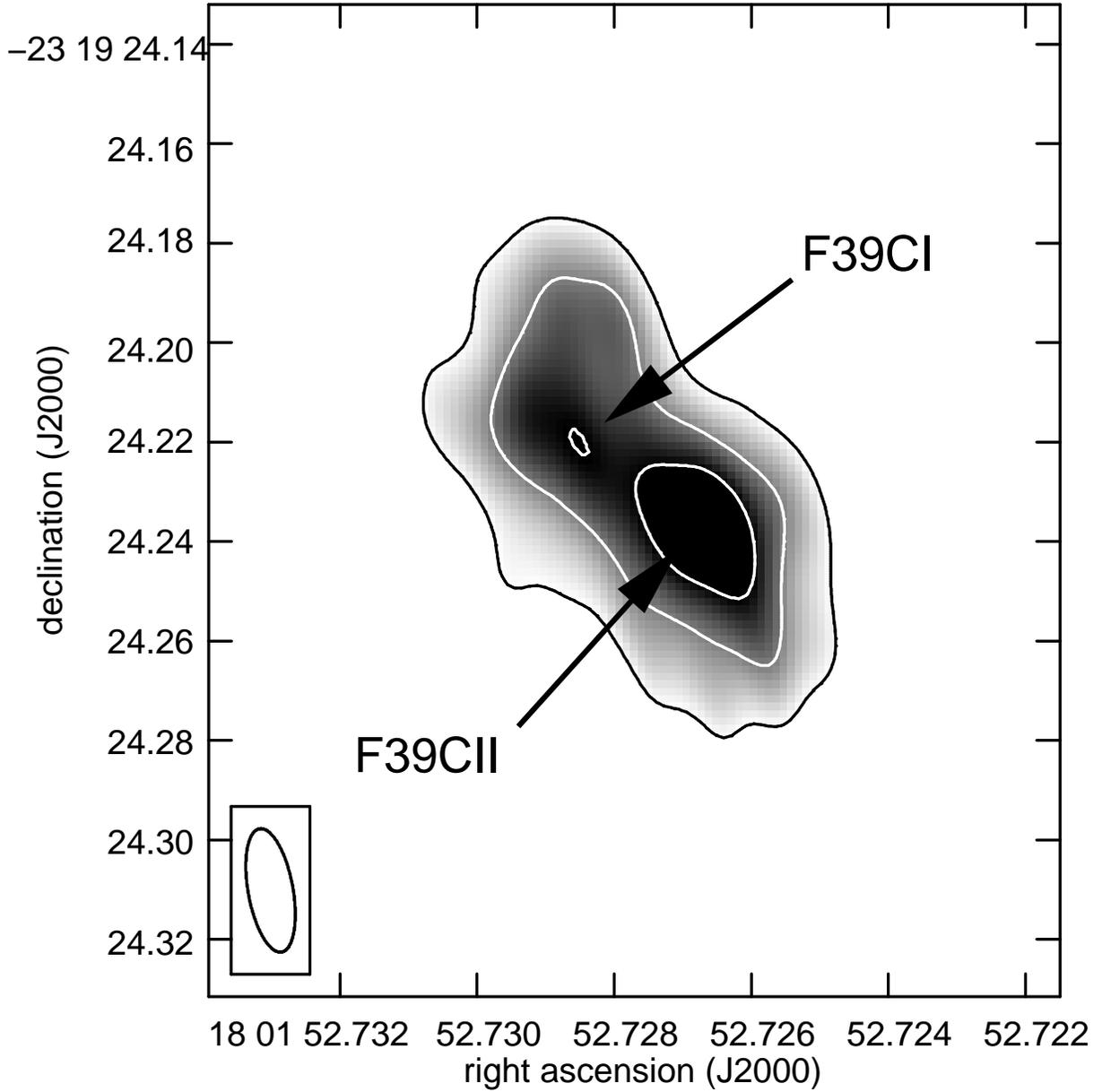}
\caption{Contour and greyscale image of the F39C maser region using the $v_{\rm LSR} = 9.7$~\kms\ channel from the VLBA observations.
The contour levels are -14, 14, 28, and 42 times the image RMS noise level of 30~\mjb\ (no negative contours appear).
The beam, plotted in the lower left-hand corner, is $25 \times 9$~mas at a position angle of 10\arcdeg.
The two image features are labelled (Table~\ref{vlbapos}).
\label{VLBA_F39C}}
\end{figure}

All data reduction and imaging was performed with the AIPS\footnote{The Astronomical Image Processing System (AIPS) is documented at {\tt http://www.nrao.edu/aips/}.} software package.
Figures~\ref{moment0}, \ref{moment0-3}, and \ref{moment0-1} show the line-integrated (zeroth-moment) MERLIN images of 19 of the masers.
The majority of maser detections are in the E and F maser regions; the MERLIN observations contain only three detections outside of regions E and F and the VLBA observations contain no detections outside of regions E and F.
These detection rates are consistent with expectations based on previous observations.
Of the 41 maser spots detected in W28 by C97 using the VLA, 23 were observed to have intensities greater than 1~\jb.
The MERLIN observations contain detections of 21 of these 23 masers; the only exceptions are the A2 and E32 masers.
Even though the A maser region lies in a sidelobe of the Lovell primary beam, the A3 maser was detected.
However, since its amplitude is very uncertain we will not discuss it further.
In addition, the MERLIN observations indicate that there are multiple angular components in the F20 and F39 masers (Figs.~\ref{moment0}, \ref{moment0-3}).
Figures \ref{VLBA_F39A} and \ref{VLBA_F39C} show the peak-channel images of the F39A and F39C masers from the VLBA+Y1 observations.

Tables \ref{mersizes} and \ref{vlbasizes} list the deconvolved angular sizes and total flux densities observed using MERLIN and the VLBA.
The F40 maser is detected on only the Y1-PT baseline (VLA-Pie Town, 52~km station separation, 700~mas fringe spacing) and Y1-LA baseline (VLA-Los Alamos, 226~km, 150~mas) in the VLBA+Y1 observations and was not imaged.
The observed size of the F40 maser in the VLBA data set is estimated to be $350\pm50$~mas from the two fringe visibilities.
Four of the 28 masers detected using MERLIN are detected using the VLBA+Y1 (E24, F39A, F39C, and F40) for a detection rate of approximately 15\%.
Using the VLBA+Y1, H03 were able to image one of the five masers in IC~443 that is detected using MERLIN for a comparable detection rate of approximately 20\%.

The deconvolved sizes of the MERLIN maser images are in the range 90 to 350~mas (225 to 875~AU projected linear size).
Compact cores with deconvolved sizes of 20~mas (50~AU) are apparent in the VLBA images.
These findings are consistent with theoretical expectations (Lockett et al.\ 1999; Wardle 1999) and previous observations (C99; H03).
Also, the total flux density of the masers as measured from the MERLIN images is in good agreement with the VLA data of C97, indicating that none of the maser emission is resolved out using MERLIN.
The VLBA images of the F39A and F39C maser emission cores contain approximately 75\% of the flux density detected in observations using the VLA and MERLIN, consistent with previous observations (C99; H03).

The {\em absolute} positions in the current VLBA data have a 3-$\sigma$ uncertainty of about 100~mas since the observations did not include an astrometric-quality phase calibrator ({\it e.g}., Reid 1999).
However, the current MERLIN observations were phase-referenced to the calibrator 1748-253 and have a 1-$\sigma$ absolute position uncertainty of about 5~mas.
The VLBA observations of C99 were also phase-referenced to 1748-253 and have a 1-$\sigma$ absolute position uncertainty of 3~mas.
The current maser positions measured using MERLIN are consistent with the positions measured by C99.
For a distance to the masers of 2.5~kpc (Vel\'{a}zquez et al.\ 2002), the agreement between the C99 observations in 1997 May and the current MERLIN observations in 2002 January implies a 3-$\sigma$ upper limit of 60~\kms\ on the transverse speed of the masers.
Draine, Roberge, \& Dalgarno (1983) suggest speeds in the range $10-50$~\kms\ ($3-16\ {\rm mas}\,{\rm yr}^{-1}$) for the shocks in which OH (1720~MHz) SNR masers occur (see H03 for a complete discussion of the potential for proper motion in OH SNR masers).
Since the current observations are not sensitive to the proper motions that may have occurred since the C99 observations (a few milliarcseconds), we adopt the C99 position for the F39A maser in this paper.

The {\em relative} positions of the masers to the F39A maser are limited only by the signal-to-noise ratio and the angular resolution of the observations, not by the initial phase calibration.
For example, the current VLBA observation of the F39C maser (40:1 signal-to-noise, 15~mas beam) has a 1-$\sigma$ relative position uncertainty of $< 1$~mas with respect to the F39A maser position.
Furthermore, the relative separation of the two main peaks in the current VLBA images of the F39A ($27$~mas, 70~AU) and F39C ($40$~mas, 100~AU) masers are consistent with the VLBA observations by C99, indicating a 1-$\sigma$ upper limit on the relative motion among the multiple maser emission cores of less than 1~mas (2.5~AU; a relative transverse velocity $<2$~\kms).

\subsection{Line Profiles}

Tables \ref{merpos} and \ref{vlbapos} list the fitted positions and deconvolved spectral-profile properties of the maser lines from the MERLIN and VLBA observations.
All line profiles were fitted using the GIPSY\footnote{The Groningen Image Processing System (GIPSY) is documented at {\tt http://www.astro.rug.nl/${\mathtt \sim}$gipsy/}.} software package.
Typical linewidths are slightly less than 1~\kms, with values measured using the VLBA being smaller (narrower) than values measured using MERLIN (the trend of observed line width decreasing with beam size is observed for many maser species, {\it e.g}.\ OH [4765~MHz]: Palmer, Goss, \& Devine 2003; H$_2$CO [4830~MHz]: Hoffman et al.\ 2003b).
This difference in line width between the MERLIN and the VLBA data is also observed for the OH (1720~MHz) masers in IC~443 (H03).
However, unlike the IC~443 observations, none of the line widths observed in W28 are below the Doppler line width of 0.5~\kms ({\it i.e}.\ in the current data $\Delta{v}/v > (2/c)(2kT\ln{2}/m)^{1/2}$, where $c$ is the speed of light, $k$ is Boltzmann's constant, $m$ is the mass of the OH molecule, and $T$ is the kinetic temperature of the maser gas, which we assume to be 90~K, as suggested in \S 1).
Also listed in the tables are the brightness temperatures $T_B$ of the masers, derived using the relation $T_B = (0.565 \lambda^2 S)/(k \Theta_a \Theta_b)$, where $S$ is the flux density of the maser (${\rm W}\,{\rm m}^{-2}\,{\rm Hz}^{-1}$), $\lambda$ is the wavelength of the radiation (m), $k$ is Boltzmann's constant, and $\Theta_a$ and $\Theta_b$ are the major and minor axes of the synthesized beam (radians) ({\it e.g}.\ Rybicki \& Lightman 1979).

\subsection{Zeeman Circular Polarization}

\subsubsection{Resolved Splitting}

The OH (1720~MHz) SNR masers in W28 were imaged separately in right-circular polarization (RCP, using RR cross-correlations) and left-circular polarization (LCP, using LL cross-correlations).
Although the RCP and LCP images of a given maser are coincident in sky position and spectral channel (the instrumental channel separation is approximately 0.2~\kms, \S2), the Gaussians fitted to the RCP and LCP line profiles have different center velocities (separated by approximately 0.07~\kms).
We assume that the relative displacement of the RCP and LCP lines is due to Zeeman level-splitting of the maser transition in a magnetic field.
Both the MERLIN and VLBA observations indicate resolved Zeeman-splitting in this manner.

The spectral line observations of the masers in this data set have both high signal-to-noise ($\sim$2500:1 for the F39A maser in the MERLIN data) and well-resolved spectral lines (the lines span about 15 spectral channels).
For these relatively high-confidence detections, the uncertainty in the measurement of the relative displacement of the fitted RCP and LCP line centers is approximately $10\ {\rm m}\,{\rm s}^{-1}$.
Also, the symmetric fit-residuals discussed in \S4.2 do not affect the confidence with which the central velocity is fitted.

The velocity difference is converted to magnetic field using the relationship $v_{\rm RCP} - v_{\rm LCP} = Z B_{\rm RL}$, where $B_{\rm RL}$ denotes the magnetic field measured using RCP and LCP line profiles and $Z$ is the Zeeman splitting coefficient ({\it e.g}.\ Fish et al.\ 2003).
The magnetic substates of the OH ground-state satellite lines have the ratio of intensities 6:3:1 ({\it e.g}.\ Davies 1974).
For thermal emission and absorption, all of these substates contribute to the observed spectrum and the Zeeman splitting factor is their weighted average $Z = 0.236\ {\rm km}\,{\rm s}^{-1}\,{\rm mG}^{-1}$ ($1.35\ {\rm Hz}\,\mu{\rm G}^{-1}$).
However, for maser amplification, the strongest of these substates is dominant ({\it e.g}.\ Gray et al.\ 1992; Caswell 2004).
For the saturated OH satellite-line masers in the current study, we adopt $Z = 0.114\ {\rm km}\,{\rm s}^{-1}\,{\rm mG}^{-1}$ ($0.654\ {\rm Hz}\,\mu{\rm G}^{-1}$).
(For an alternate viewpoint, compare the discussion in Fish et al.\ [2003] concerning their choice of $Z = 0.236\ {\rm km}\,{\rm s}^{-1}\,{\rm mG}^{-1}$.)
Table~\ref{circ} lists the velocity differences and corresponding magnetic fields for the masers in which the Zeeman splitting is resolved.

\subsubsection{Stokes-$V$ Fitting}

The results presented in \S3.3.1 represent the first observation of resolved Zeeman splitting in OH (1720~MHz) masers.
Although these resolved-Zeeman-splitting results already provide complete information about the magnetic field strength, we present the following Stokes V fitting results as well since previous Zeeman analyses of OH (1720~MHz) SNR masers have employed the fitting method described in this section.

Zeeman analyses are based upon the degree to which the Zeeman components have been separated in frequency from the resonance frequency of the line transition.
The separation may be parameterized with $x_B \equiv \Delta{\nu_B}/\Delta{\nu_D}$, where $\Delta{\nu_B}$ is the Zeeman frequency separation of the split substates and $\Delta{\nu_D}$ is the Doppler-broadened width of the maser line, typically the same as the width of the saturated Stokes $I$ maser line profile (cf.\ H03; Watson \& Wyld 2003).
The Zeeman splitting in the current data set is $x_B \approx 0.1$ (Tables \ref{merpos}, \ref{vlbapos}, \ref{circ}).
This degree of splitting may violate the assumption ($x_B \ll 1$) upon which most theoretical investigations are based ({\it e.g}.\ Goldreich, Keeley, \& Kwan 1973), as will be discussed in \S\ref{circ_disc}.

In the regime $x_B \ll 1$, the following equation relates the Stokes $I$ and Stokes $V$ profiles ({\it e.g}.\ Roberts et al.\ 1993; Sarma et al.\ 2001; BFGT),
\begin{equation}
V = b \frac{\partial{I}}{\partial{\nu}} + aI\ .
\label{zeemap}
\end{equation}
The Stokes $I$ and $V$ profiles of the maser lines from both the MERLIN and VLBA observations were fitted to equation~\ref{zeemap}.
All fitting was performed with the MIRIAD\footnote{The Multichannel Image Reconstruction, Image Analysis and Display (MIRIAD) is documented at {\tt http://bima.astro.umd.edu/miriad/}.} software package.

The $aI$ term in equation~\ref{zeemap} represents the symmetrical component of the Stokes $V$ profile.
The fitted sign and magnitude of $a$ is consistent with instrumental ``leakage'' of Stokes $I$ into Stokes $V$ ({\it e.g}.\ Crutcher et al.\ 1975); in the current data $a$ does not represent an intrinsic property of the maser radiation.
For example, the fits to the MERLIN data for the F39A and E24 masers yield $a = 0.4$\% and $a = 0.2$\%, respectively, consistent with the expected instrumental leakage (0.3\%) of the MERLIN array.\footnote{see the MERLIN User Guide, {\tt http://www.merlin.ac.uk/user\_guide/OnlineMUG/} }
However, we investigated whether $a$ may represent an intrinsic property of the maser emission.
In the case of unsaturated amplification, a symmetric component of Stokes $V$ is expected ({\it e.g}.\ Vlemmings, Diamond, \& van Langevelde 2002) but must be accompanied by a narrowing of $\partial{I}/\partial{\nu}$ with respect to Stokes $V$, an effect which is not observed in the current data (also see BFGT).
Therefore, we have subtracted the $aI$ term from the data and do not consider it in the current analysis.

The conversion between the fit parameter $b$ and the magnetic field strength has differing theoretical interpretations, to be discussed in \S\ref{circ_disc}.
Since many of the relevant current theories suggest that $b$ depends on $\theta$, the angle between the line of sight and the magnetic field, we denote the magnetic field strength determined from equation \ref{zeemap} as $B_\theta$, to distinguish it from $B_{RL}$ which is measured as described in \S3.3.1.
In this paper the values of the fitted parameter $b$ are related to the magnetic field $B_\theta$ using $b = Z B_\theta/2$.
Note that the factor of $1/2$ is associated with the derivative of Stokes $I$ since $I=2R$ or $2L$ (in the case of equal amplification of both polarizations) and $V=R-L$.
From Table 6 it is clear that the derived values of $B_{RL}$ and $B_\theta$ are in good agreement.
This correspondence will be discussed further in \S4.3.

The improvement in the spectral and angular resolution from the C97 VLA observations to the MERLIN observations has allowed a more reliable determination of the polarization properties of the line profiles.
For example, the F20B and E27 masers are shown to have multiple spectral components based on the MERLIN observations (Table \ref{merpos}).
Furthermore, the C97 VLA experiment yielded only 11 Zeeman detections from 41 image features while the current MERLIN observations yield 23 Zeeman detections from 28 image features (Table \ref{circ}).
Low signal to noise, not confusion due to blending, prevented Zeeman fitting of five of the 28 masers observed using MERLIN.

We can directly compare the derived magnetic field strengths from this work with those of C99.
Such a comparison shows that the field strengths reported in C99 are systematically higher by about a factor of 4.
We do not understand this discrepancy.
Indeed, we have re-analyzed the data of C99 in exactly the same manner as we analyzed the new data presented here.
In doing so, we obtain comparable results for the strength of the magnetic field fitted to both the the C99 data and the current data.
For example, in our re-analysis, the magnetic field derived for the position marked F39B using the C99 data (which is labeled F39AI in the current Figure 5) is $0.41 \pm 0.06$~mG.
This compares favorably with the field determined from the current data as $0.31 \pm 0.02$~mG.
Given the similar results from the re-analysis, we suggest that C99 had a conversion error in the calculation of their field strengths.

\subsection{Linear Polarization}

\begin{figure}
\plotone{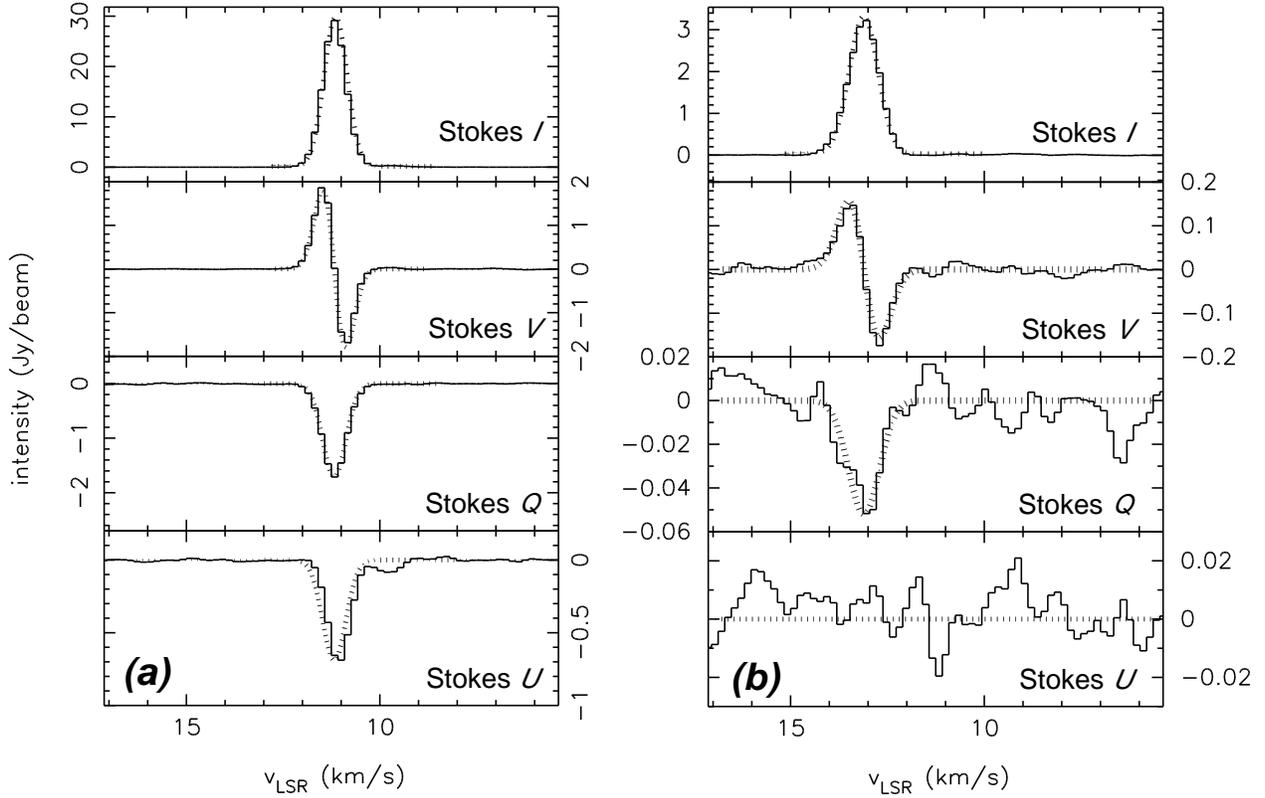}
\caption{Typical full-polarization spectra from the MERLIN observations for the {\it (a)} F39A and {\it (b)} E24 masers.
The dashed lines are, from top panel to bottom panel, the best-fit Gaussian to Stokes $I$, the derivative of the best-fit Gaussian scaled with the fitted Zeeman magnetic field, and the best-fit Gaussian scaled with the Stokes $Q$ and $U$ amplitudes listed in Table~\ref{merpol}.
\label{4pol}}
\end{figure}

\begin{figure}
\plotone{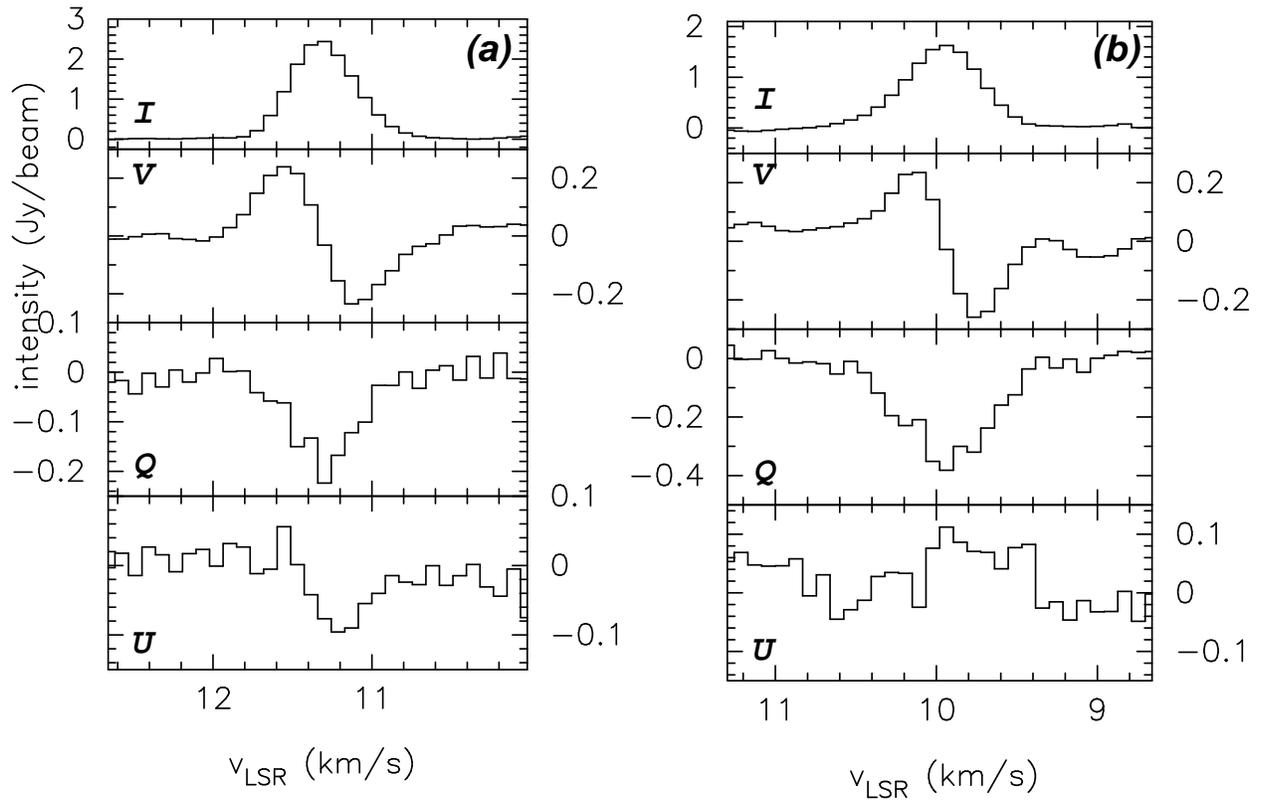}
\caption{Full-polarization spectra from the VLBA observations for the {\it (a)} F39AI and {\it (b)} F39CII masers (Table~\ref{vlbapol}).
\label{vlba4pol}}
\end{figure}

The linear polarization angles $\chi$ listed in Tables \ref{merpol} and \ref{vlbapol} are determined using the usual relation $\tan{(2\chi)} = (U/Q)$, where $Q$ and $U$ are Stokes parameters calculated from the RL and LR cross-correlations (\S 2).
The polarization angles are in good agreement with those observed by C97 using the VLA.
Figure \ref{4pol} shows typical full-polarization profiles for two of the masers observed using MERLIN.
The dotted lines in Figure \ref{4pol} also show the results of our Stokes $I$ Gaussian analysis, Stokes $V$ fit using equation \ref{zeemap}, and the Stokes $I$ fit scaled to the appropriate value to match the amplitudes of $Q$ and $U$ listed in Table \ref{merpol}.
Figure \ref{vlba4pol} shows two full-polarization spectra from the VLBA observations.
The percentage of linear polarization found for the W28 OH (1720~MHz) masers ranges from approximately 5\% to 20\%.
The exact values for each maser are also listed in Tables \ref{merpol} (MERLIN) and \ref{vlbapol} (VLBA).

\begin{figure}
\plotone{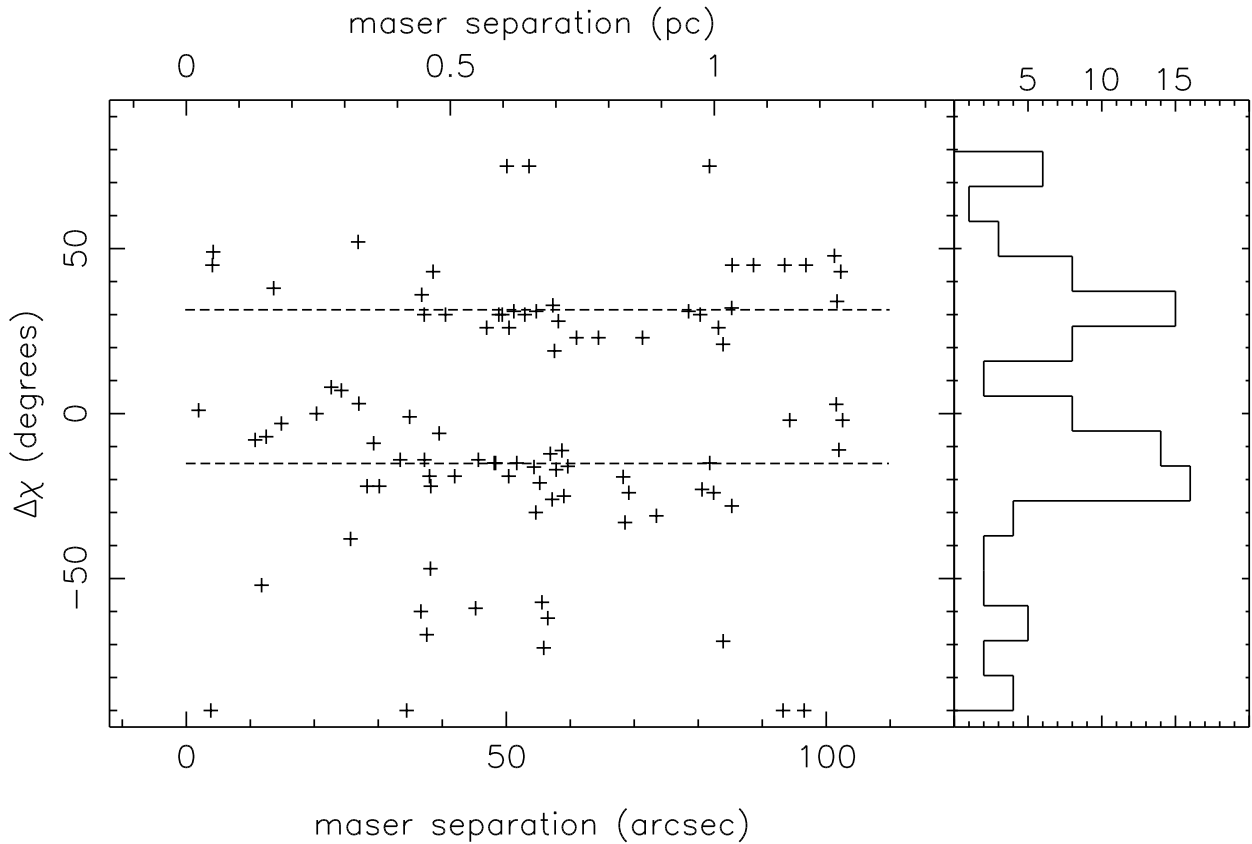}
\caption{{\it (a)} The difference in linear polarization position angle $\Delta{\chi}$ observed using MERLIN (Table~\ref{merpol}) between pairs of masers as a function of their separation (see \S 4.4).
The order of the subtraction in the difference follows increasing right ascension ({\it i.e}.\ $\Delta{\chi}$ = [$\chi$ of maser with lower $\alpha$] - [$\chi$ of maser with higher $\alpha$]).
The dashed lines, which have a slope of zero, are plotted in order to illustrate that $\Delta{\chi}$ is relatively independent of the separation of the masers.
{\it (b)} A histogram of $\Delta{\chi}$ summed over the relative separation.
\label{deltaChi}}
\end{figure}

For every pair of masers in the MERLIN observations, the difference in polarization angles $\Delta{\chi}$ has been plotted in Figure \ref{deltaChi} as a function the relative separation of the masers.
The histogram in the figure indicates that the majority of the masers (approximately 55\%) have $\Delta{\chi} \approx 0\arcdeg$, with a secondary population (approximately 30\%) near $\Delta{\chi} = 45\arcdeg$.
The dashed lines in Figure \ref{deltaChi} have a slope of zero, indicating that the difference in polarization angles is predominantly independent of maser separation.
Thus, the masers have comparable polarization angles throughout the approximately 2 arcminutes spanned by the E and F maser regions.

\begin{figure}
\plotone{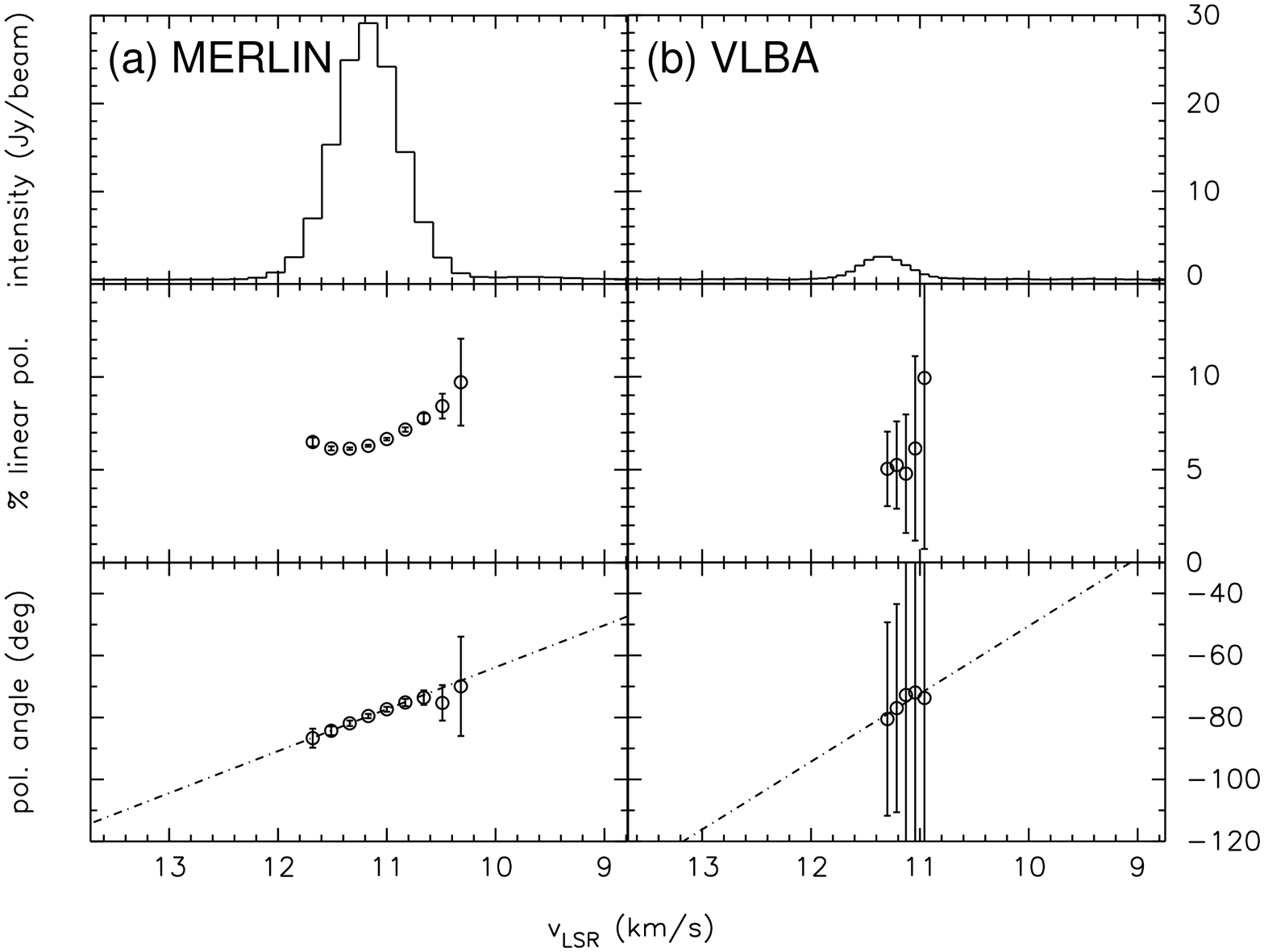}
\caption{Comparison of the highest signal-to-noise detections (F39A) for the {\it (a)} MERLIN and {\it (b)} VLBA observations, showing {\it (top)} the total intensity spectral line profile, {\it (center)} the fraction of linearly polarized emission ($q$), and {\it (bottom)} the $\partial{\chi}/\partial{\nu}$ profile.
The dashed line is the best-fit linear $\partial{\chi}/\partial{\nu}$ slope discussed in \S \ref{lin_prof_disc}.
\label{dxdv}}
\end{figure}

\begin{figure}
\plotone{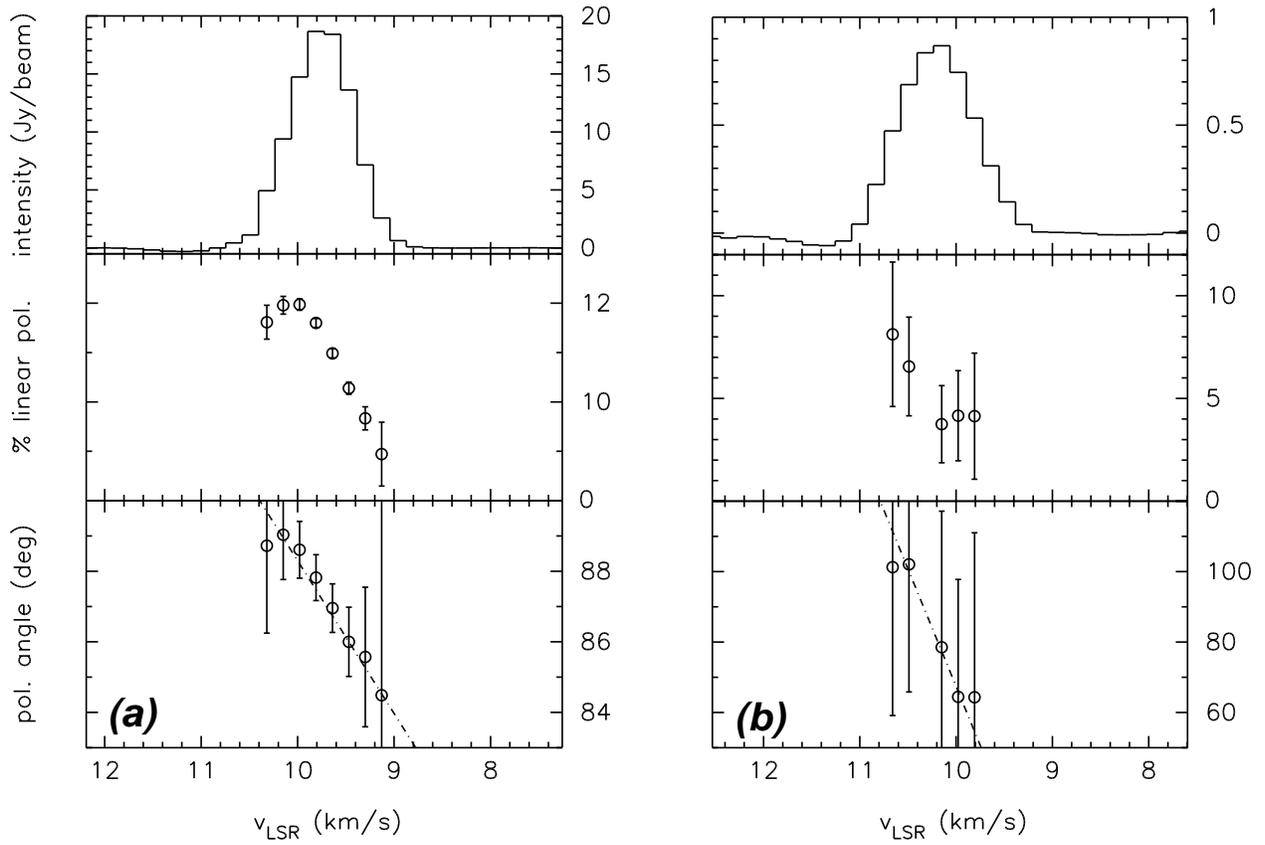}
\caption{Like Figure~\ref{dxdv}, showing the $\partial{\chi}/\partial{\nu}$ and $q$-profiles (see \S \ref{lin_prof_disc}), for the MERLIN observations of the {\it (a)} F39C and {\it (b)} F21 masers.
\label{2pol}}
\end{figure}

The value for $\partial{\chi}/\partial{\nu}$ listed in Tables \ref{merpol} and \ref{vlbapol} is the observed change in polarization angle across the maser line profile.
Figures \ref{dxdv} and \ref{2pol} show that $\partial{\chi}/\partial{\nu}$ is approximately linear.
This result is discussed in \S \ref{lin_prof_disc}.
The values of $\chi$ listed in the tables and figures are the values at the center of the line.

\section{Discussion}

As described in \S1, previous OH (1720~MHz) SNR masers studies were based on arcsecond-resolution observations with instruments such as the VLA ({\it e.g}.\ BFGT).
In the current paper, we describe MERLIN and VLBA observations which have more favorable angular and spectral resolution.
As expected, these new observations have sufficiently alleviated instrumental blending to permit reliable measurements of the intrinsic emission properties of the masers.
However, in interpreting these results in this section of the paper, we must rely on the existing literature concerning theoretical investigations of the maser emission.

The theoretical investigations upon which our analyses are based may be grouped into three topics, listed here in order of decreasing consensus:
(1) the population inversion of the 1720-MHz transition,
(2) the degree of saturation of the maser gain process, and
(3) the polarization of the maser emission due to the presence of a magnetic field in the maser gas.

In \S4.1, we discuss the relative location, size, and brightness of the W28 masers.
In the ten years since the rediscovery of OH SNR masers (Frail, Goss, \& Slysh 1994), there has been a successful convergence between observation and theory concerning the expected size and excitation process of OH SNR masers (H03; Claussen et al.\ 2002; Lockett et al.\ 1999; Wardle 1999).
In this paper, we add further evidence of the agreement between theory and observation (\S3.1) and describe some new results pertinent to the milliarcsecond scale of the current observations.

In \S4.2, we discuss the degree to which the masers are saturated.
Using two independent tests, we find that the masers must be at least partially saturated.
However, many of the relevant maser polarization theories depend sensitively on the exact degree of maser saturation, which we are unable to accurately constrain.

In \S\S4.3, 4.4, and 4.5, we compare the observed polarization properties with existing maser theories, including new work since this this topic was reviewed by BFGT (e.g. WW01; Gray 2003).
However, we emphasize that all of these theories employ two limits: either $x_B \ll 1$ (typical of the thermal Zeeman case) or $x_B > 1$ (fully resolved Zeeman splitting typical of masers found in star forming regions) that may render them inapplicable to our specific case where $x_B \approx 0.1$.
For the time being we are stuck with this ambiguity, but hope that these observational data will spark interest in a theoretical study appropriate to the SNR OH (1720 MHz) maser regime.
Thus, the ``$x_B$ limit'' caveat imposed by the currently available theories should be kept in mind throughout these sections.

\subsection{Angular Morphology}

Observations which span a wide range of spatial scales suggest that the OH (1720 MHz) masers exhibit both (1) a large region (500-1000~AU) of relatively weak emission surrounding, in some cases, (2) multiple, closely-spaced (on the order of 100~AU), compact cores (50-100~AU) of emission (C99; H03).
In general, radio interferometers such as the VLA (baseline lengths on the order of 10~km), MERLIN ($\sim 100$~km), and VLBA ($\sim 1000$~km) are sensitive to different size-scale regimes of this emission morphology.

The W28 SNR, and OH (1720~MHz) SNR's in general, contain several physically distinct maser regions (discerned using differing velocities, polarization properties, etc.) separated by angular distances of a few hundred milliarcseconds to several arcminutes.
Observations using the VLA (with beam sizes $\geq 1\arcsec$) yield images in which several emission regions are blended into single image spots.
Observations using MERLIN (with beam sizes of approximately 100~mas) yield images in which every image feature represents a physically distinct region of maser emission (Tables \ref{mersizes}, \ref{merpos}, \ref{merpol}).
For example, Figure~\ref{moment0} shows four masers in the F39 region with distinct velocities, Zeeman profiles, and linear polarization properties (Tables \ref{circ} \& \ref{merpol}) at the position where observations using the VLA indicate only one image feature for which the emission properties could not be determined (C97).

Observations at the position of a given MERLIN maser image using the VLBA (approximately 10-milliarcsecond angular resolution) reveal multiple image peaks, just as MERLIN observations of a VLA source reveal several image peaks.
However, these image peaks observed using the VLBA {\em do not} have different velocities (on the scale of a line width) nor different polarization properties (see also C99; H03).
Indeed, the multiple peaks in the VLBA images are more likely due to inhomogeneities within a single region of maser emission than to several physically distinct maser regions coincident along the line of sight.
For example, MERLIN observations at the position of F39 contain the F39A and F39C masers (Fig.~\ref{moment0}).
The velocities of the F39A and F39C masers are $v_{\rm LSR} \simeq 11.2$~\kms\ and $v_{\rm LSR} \simeq 9.7$~\kms, respectively (Table \ref{merpos}).
These velocities are separated by a few line widths and indicate a difference in systemic kinematics between the F39A and F39C regions.
In contrast, the peaks observed in the VLBA images (Fig.~\ref{VLBA_F39A}, \ref{VLBA_F39C}), F39AI--III and F39CI--II (Table \ref{vlbapos}) are separated by much less than one line width in velocity (a separation of approximately 0.04~\kms\ compared with a line width of approximately 0.60~\kms).
Therefore, we conclude that the different peaks in the VLBA images of OH (1720~MHz) SNR masers are not dynamically distinct regions of maser gas, but are instead emission inhomogeneities within a single volume of maser gas.

\subsection{Saturation of the Maser Amplification}

A critical element in the analysis of maser radiation is the degree to which the maser is saturated ({\it e.g}.\ Vlemmings et al.\ 2002; WW01; Elitzur 1998).
For the case of OH (1720~MHz) SNR masers we discuss two tests: degree of linear polarization and line profile shape.

In order for maser emission to possess linear polarization, it is expected on theoretical grounds ({\it e.g}.\ Nedoluha \& Watson 1990) that the maser needs to be at least partially saturated.
This expectation has been used as evidence for partial saturation in the case of the OH (1720~MHz) SNR masers in W28, W44, IC443, and W51, which all show linear polarization as observed using the VLA (C97; BFGT).
In this paper, the MERLIN observations contain 22 detections of linear polarization.
Furthermore, there are no non-detections of linear polarization with confidence higher than 3-$\sigma$.
Thus it seems likely that the OH (1720~MHz) SNR masers are at least partially saturated.

\begin{figure}
\plotone{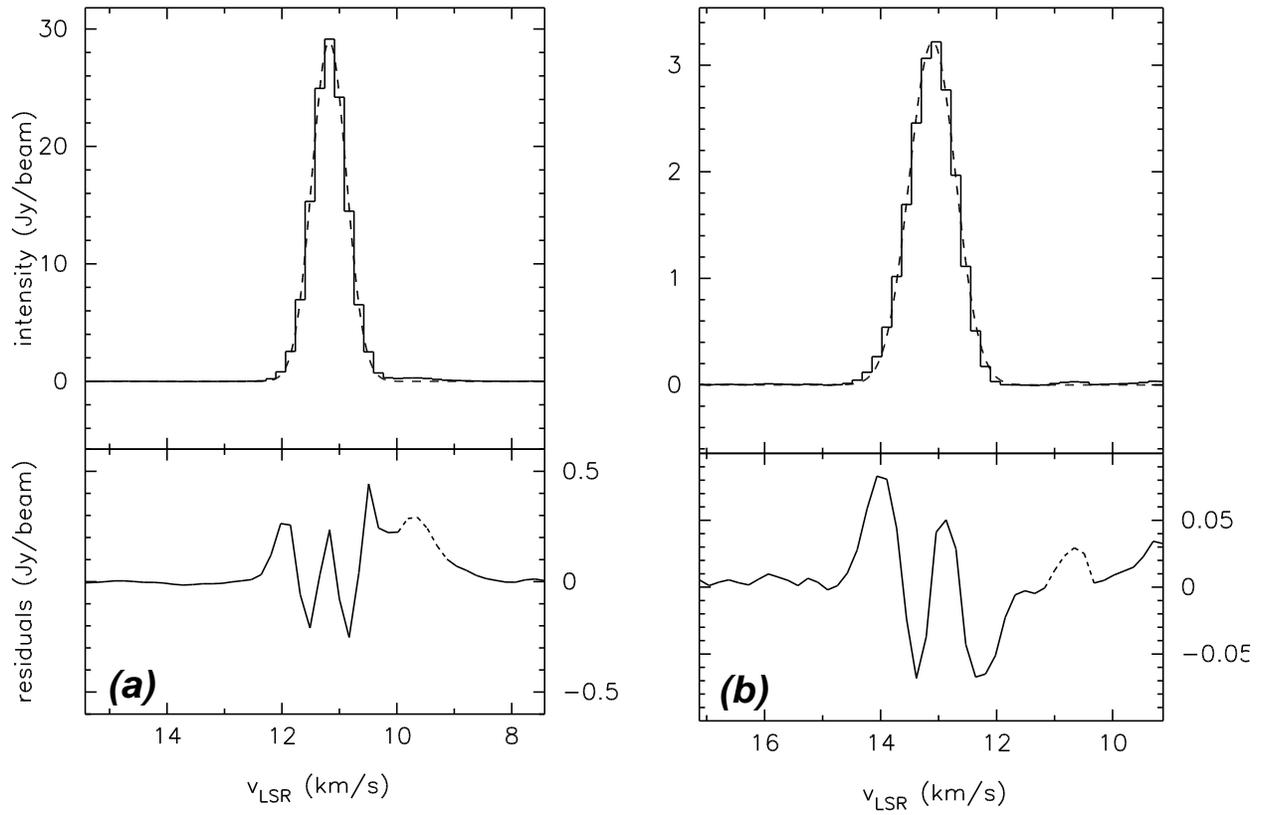}
\caption{Best-fit Gaussians and fit residuals for the MERLIN profiles of the {\it (a)} F39A and {\it (b)} E24 masers.
The portions of the residuals plotted using a dotted line are due to imaging artifacts in off-peak channels which are not included in the fitting or in the analysis.
\label{resid}}
\end{figure}

Another test results from the fact that unsaturated maser emission is expected to deviate substantially from a pure Gaussian profile (see, for example, Elitzur 1998).
Watson, Sarma, \& Singleton (2002) have examined the deviation of maser line shapes from Gaussian.
Figure~\ref{resid} shows the fitted Gaussians and residuals to the E24 and F39A maser lines from the MERLIN data.
In the Watson et al.\ model, the Gaussian deviation of the line profile is quantified using the two parameters $\delta$ (defined in Watson et al.\ 2002) and kurtosis (the normalized fourth central moment of a distribution: {\it e.g}.\ Abramowitz \& Stegun 1972).
The fit residuals of the line profiles observed using MERLIN have deviation parameters similar to the H$_2$O masers considered by Watson et al.\ (2002; $\delta \approx 3 \times 10^{-4}$ and a negative kurtosis), indicating that the OH (1720~MHz) SNR masers are partially saturated (have a normalized model intensity near unity).

Both of these tests indicate at least partial saturation of the OH (1720~MHz) SNR masers.
Therefore, the saturated regime of the maser polarization theories proposed by both Watson et al.\ and Elitzur will be applied in the following discussion.

\subsection{Zeeman Detections\label{circ_disc}}

The magnetic fields of the W28 masers have been estimated using two techniques: resolved-Zeeman splitting and fitting to Stokes $V$.
Magnetic fields calculated directly from resolved Zeeman splitting ($B_{RL}$) are generally thought not to depend on $\theta$, and thus this technique measures the full magnetic field strength ({\it e.g}.\ Townes \& Schawlow 1955).
The fitting of the Stokes $V$ profiles to obtain $b$ using equation \ref{zeemap} (\S3.3.2) is typically used for measuring the Zeeman effect in thermal gas.
In the thermal case, $b$ is proportional to $B\cos{\theta}$, but as mentioned in \S3.3.2 the dependence of $b$ on $\theta$ for the maser case is uncertain.
For the first time we are able to directly compare these two techniques.

For the maser case where $x_B \ll 1$, a number of studies have examined the dependence of $b$ on $\theta$ for maser emission.
Nedoluha \& Watson (1992) and Watson (2002) suggest that for unsaturated masers, the thermal interpretation ({\it i.e}.\ $b\propto B\cos{\theta}$) is appropriate.
However, as described in \S4.2, the SNR OH (1720~MHz) masers are likely at least partially saturated.
In the case of a saturated maser, the $\theta$ dependence is both more controversial and complex; possibly depending on a geometrical term in addition to $\theta$.
Here we only concentrate on the $\theta$ dependence.
Elitzur (1998) suggests that $b \propto \cos{\theta}^{-1}$, while WW01 present numerical simulations that yield different results (although without a mathematical expression it is difficult to express exactly how different).
Alternatively, $b$ may be independent of $\theta$ in the $x_B \approx 0.1$ case observed for the OH (1720~MHz) masers (M.\ Elitzur 2003, private communication), in analogy with the $x_B > 1$ maser case.

Figure 16 shows a plot of the derived $B_{RL}$ versus $B_\theta$ values for each maser.
It is clear from this plot and Table 6 that $B_{RL} \approx B_\theta$ to a high degree of coincidence.
This strong correlation leads to only two viable alternatives for the dependence of $b$ on $\theta$ for the SNR OH (1720~MHz) masers:
(1) There is an as yet undiscovered $\theta$ dependence on $b$ measured from the resolved-Zeeman-fitting case ($B_{RL}$), such that this dependence is the same as that of the Stokes-$V$ fitting case (whatever it may be); or
(2) For the $x_B \approx 0.1$ case applicable to SNR OH (1720~MHz) masers there is no appreciable $\theta$ dependence on $b$.
Although we cannot distinguish between the two, we suspect that option (2) is more likely.

Also, it has been observed (Brogan et al.\ 2003) that the magnetic fields in OH (1720~MHz) SNR masers measured using Stokes fitting depend upon the angular resolution of the observations: smaller beam areas yield larger magnetic field measurements.
This trend is also observed in other maser species ({\it e.g}.\ H$_2$O, Sarma et al.\ 2001).
The OH (1720~MHz) SNR masers in W28 show an increase in fitted magnetic field between VLA (C97) and MERLIN observations due to blending at VLA scales.
However, the magnetic field strengths fitted to the current MERLIN and VLBA observations are in good agreement.
That is, it appears that the MERLIN beam is sufficiently small to alleviate all suppression due to blending.
Indeed, since the image features in the VLBA observations have comparable emission characteristics (\S 4.1), it is not expected that an angular convolution of a VLBA image using the MERLIN beam would result in a suppression of the fitted spectral parameters (see Sarma et al.\ 2001).

\subsection{Linear Polarization\label{lin_disc}}

The linear polarization state of the masers is suggested to be a diagnostic of the orientation of the magnetic field in the maser gas.
In this section, both (1) the linear polarization fraction $q$ of the masers and (2) the orientation of the linear polarization angle $\chi$ of the masers with respect to the surrounding magnetic field are discussed.
Unlike Zeeman splitting, in which the observed circular polarizations are affected by only the line-of-sight component of the magnetic field, the polarization angle $\chi$ of the maser radiation depends on the plane-of-the-sky component of the magnetic field.
Indeed, Elitzur (1996) and Watson et al.\ (WW01) suggest that $q$ and $\chi$ depend upon $\theta$, the angle between the line of sight and the magnetic field (see also a comparison of these models by Gray [2003]).
In the case of W28, the orientation of the magnetic field in the maser region may be independently constrained using existing observations of the synchrotron continuum radiation from the SNR.
Furthermore, since the magnetic field is likely to be aligned parallel to the shocked SNR/molecular cloud interface which creates the maser conditions ({\it e.g}.\ Balsara et al.\ 2001), a comparison between the polarization state of the masers and observations of the shocked environment in W28 is also possible.

\subsubsection{Correction for Faraday Effects\label{RM}}

The linear polarization properties of the synchrotron emission from W28 on large angular scales (2\arcmin\ to 6\arcmin) have been studied by a number of groups (Milne \& Wilson 1971; Kundu \& Velusamy 1972; Milne \& Dickel 1975; Dickel \& Milne 1976; Milne 1987; Milne 1990).
These studies find that the position angle of the synchrotron linear polarization $\chi^{synch}$ is quite constant at approximately 80\arcdeg along the eastern side of the remnant where the Region E and F masers are located.
Since $\chi^{synch}$ is perpendicular to the plane of sky orientation of the magnetic field ({\it e.g}.\ Rybicki \& Lightman 1979), $PA_B^{synch}=-10\arcdeg$.
The assumption that $\chi^{synch}$ remains constant down to the size scales relevant to the maser emission of a few hundred milliarcseconds is essential to compare the two since no higher resolution synchrotron polarization studies exist.
Certainly on a few arcminute size scales the $\chi^{synch}$ is quite uniform (see for example Dickel \& Milne 1976), and also the $\chi^{maser}$ is fairly uniform from maser to maser ({\it i.e}.\ Figures 13 and 14).
However, that this is an assumption should be borne in mind during the following discussion.

For all measurements of linear polarization percentage and position angle, Faraday depolarization and rotation are of concern.
Due to the continuous nature of synchrotron emission, the above studies have been able to use the observed wavelength dependence of the polarization properties to correct for Faraday rotation and estimate the depolarization.
In particular, Dickel \& Milne (1976) have made a careful study of Faraday effects toward W28 (albeit at $2\arcmin$ sizescales).
These authors do not find evidence for Faraday rotation near the maser locations where the rotation measure is $RM \approx 0\,{\rm rad}\,{\rm m}^{-2}$.
At nearby locations, not coincident with the masers (approximately 2\arcmin\ to the east and west) the $RM$ rises to $33\,{\rm rad}\,{\rm m}^{-2}$.
Thus, we view $33\,{\rm rad}\,{\rm m}^{-2}$ as a strict upper limit, but will assume $RM \approx 0\,{\rm rad}\,{\rm m}^{-2}$ for the remaining discussion.
This assumption is limited by the fact that a value of $RM$ in the direction of the masers in excess of approximately $15\ {\rm rad}\,{\rm m}^{-2}$ (equivalent to a Faraday rotation of the position angle by approximately 25\arcdeg) could invalidate the alignment comparisons made between the synchrotron and maser polarization which we base on $RM = 0 \pm 15\ {\rm rad}\,{\rm m}^{-2}$.

As discussed in BFGT stimulated maser emission can only be Faraday depolarized along the gain length of the maser.
The gain lengths and ionization fractions suggested for SNR OH (1720~MHz) masers are such that this effect must be small (see BFGT for details).
Indeed, Dickel \& Milne (1976) find that the synchrotron depolarization toward the E and F maser locations is only 3\%.
These two facts combined suggest that there is little Faraday depolarization of the maser radiation.

\subsubsection{Interpretation of Maser Emission}

\begin{figure}
\plotone{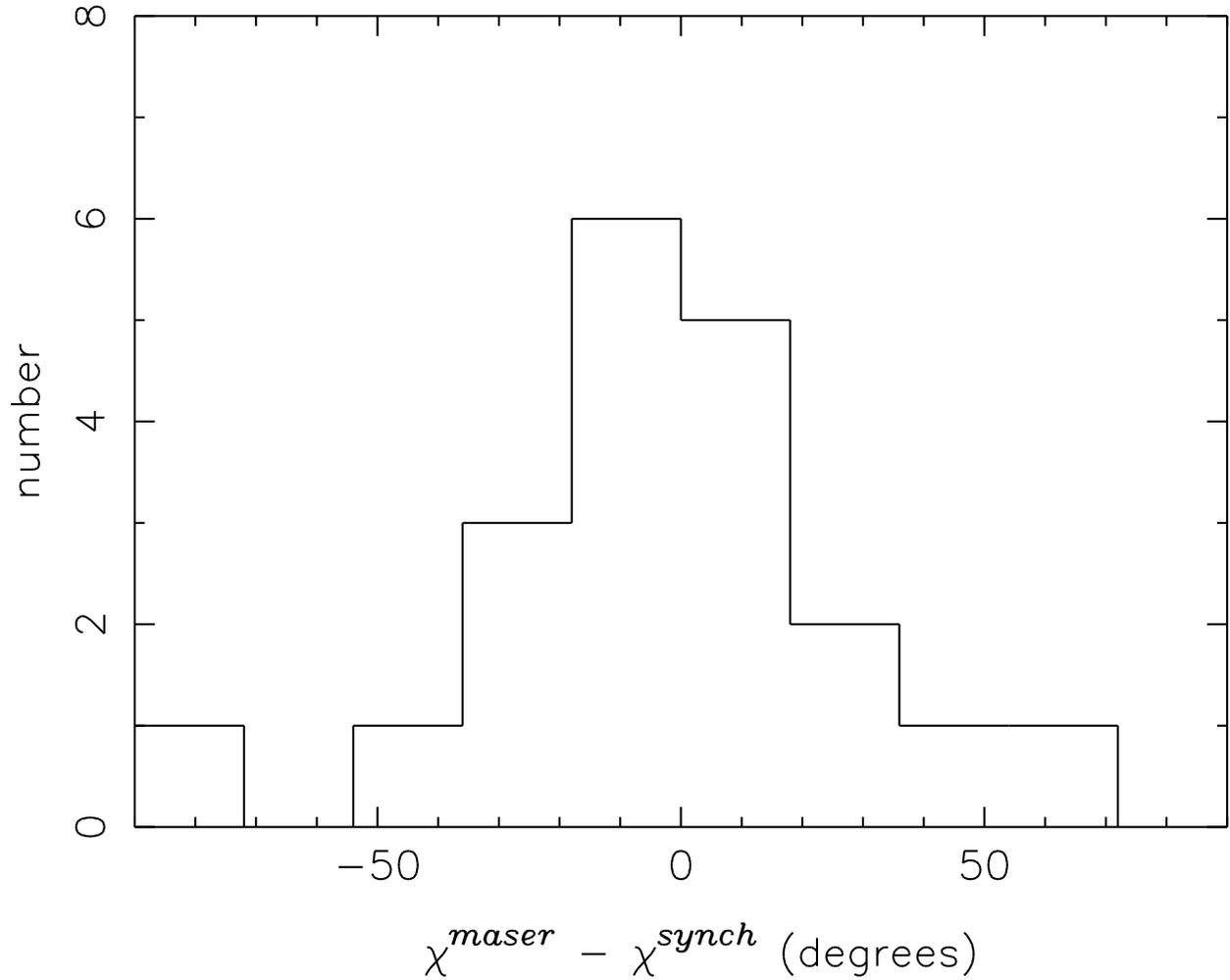}
\caption{A histogram of the difference between the polarization angle of the W28 synchrotron emission ({\it e.g}.\ Milne 1990) and individual E \& F region OH (1720~MHz) SNR maser emission from the MERLIN observations (Table~\ref{merpol}).
The polarization angle of the synchrotron emission is $\chi^{synch} = 80\arcdeg$ throughout the region.
\label{histo}}
\end{figure}

The difference between the linear-polarization angle of the synchrotron observations ($\chi^{synch}$) and the linear-polarization angles measured from the masers ($\chi^{maser}$, $\chi$ in Table~\ref{merpol}) is shown in Figure~\ref{histo}.
The difference between $\chi^{maser}$ and $\chi^{synch}$ is distributed about 0\arcdeg, with FWHM approximately 30\arcdeg, indicating a good agreement between the synchrotron and maser polarization states.
The simultaneous presence of a weaker (by about an order of magnitude), turbulent component of the magnetic field ({\it e.g}.\ Balsara et al.\ 2001) could account for the small (10\%) relative differences in the masers' polarization results (Tables \ref{merpol}, \ref{vlbapol}).
The good agreement between the polarization states of the synchrotron emission and the maser emission indicates that the maser polarization angle, $\chi$, is perpendicular to the projected magnetic field orientation (${\rm PA}_B$) in the maser region.

For the case in which $\chi^{maser} \perp {\rm PA}_B$ for maser radiation, WW01 suggest $\theta \gtrsim 55\arcdeg$ and $q \approx 5$\% for partial saturation of the maser, which is consistent with suggestions made by Elitzur (1998).
BFGT review the observations of linear polarization in the W28, W44, W51, and IC~443 masers using the VLA, which indicate polarization fractions $q \approx $ 5 to 10\%, implying $\theta \approx 60\arcdeg$.
Observations of linear polarization in the IC~443 masers using the VLBA indicate $q < 15$\% (Hoffman \& Goss, unpublished).
The current MERLIN and VLBA data (Figs.~\ref{dxdv}, \ref{2pol}) are in good agreement with VLA observations of W28 (C97) and also indicate $q \approx $ 5 to 10\% and $\theta \approx 60\arcdeg$.
From this comparison between observations and theoretical suggestions, it appears that the OH (1720~MHz) SNR masers in W28, W44, W51, and IC~443 all have $\theta \approx 60\arcdeg$.
As noted by BFGT, this is an uncomfortable level of coincidence in the absence of other constraints on the preferred direction, and may be indicative of the use of inapplicable theoretical models in our interpretation.
We plan to present a more detailed exploration of this issue in a future paper.

\begin{figure}
\plotone{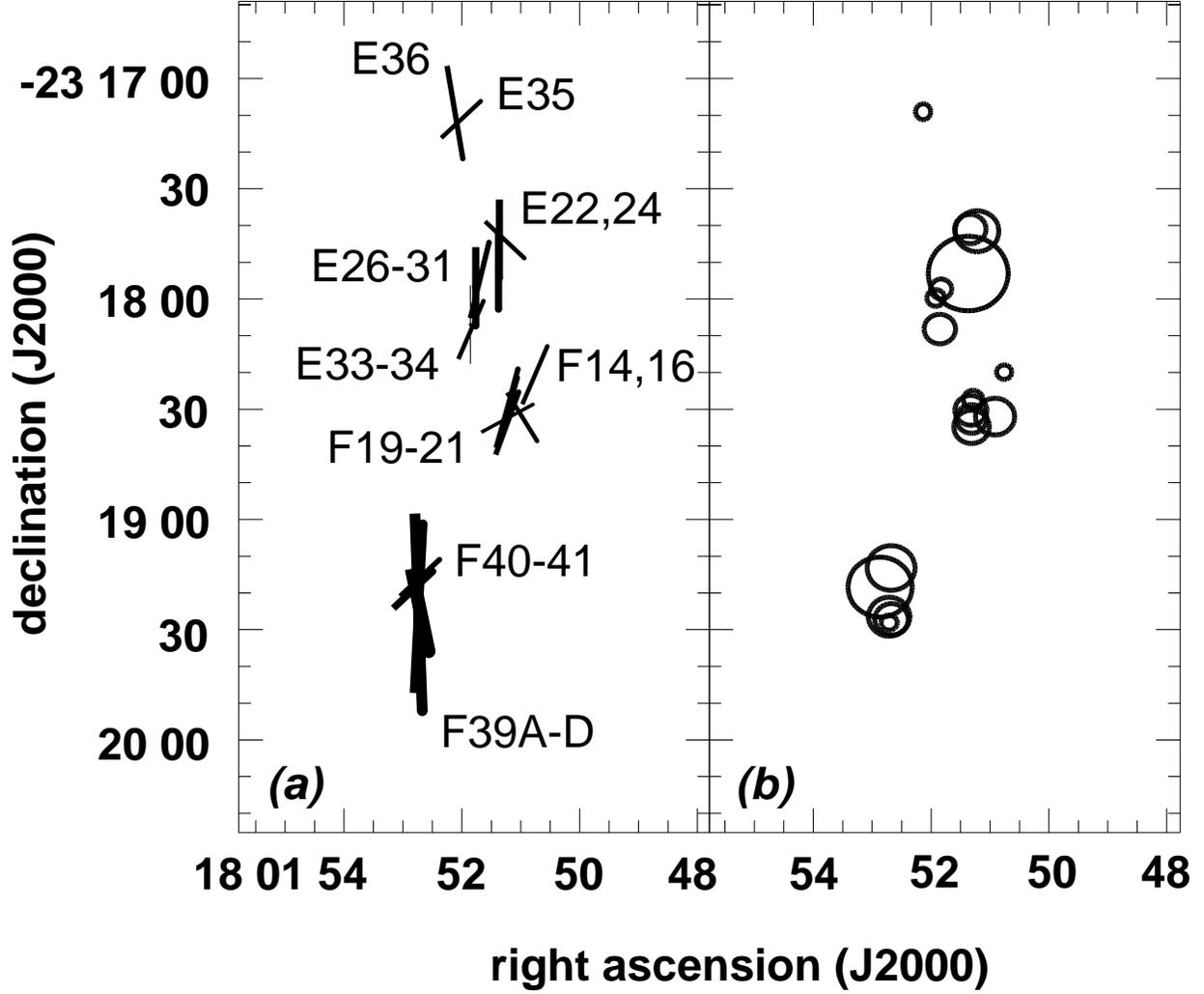}
\caption{MERLIN magnetic field results for the E and F region masers.
({\it a}) Position angle of the magnetic field (using ${\rm PA}_B^{maser} \perp \chi^{maser}$) represented with the angle of the line at the location of the maser and ({\it b}) magnitude of $B_\theta$ fitted to circular polarization Zeeman profiles represented with the diameter of the circle at the location of the maser.
These values are inferred from the maser observations as  discussed in \S 4; position-angle measurements are sensitive to the plane-of-the-sky component of the magnetic field and that Zeeman circular-polarization measurements are sensitive to only the line-of-sight component of the magnetic field.
The thickness of the lines is inversely proportional to the percent error in the measurement.
\label{pa+90}}
\end{figure}

Accepting the theoretical framework used to interpret these observations, a comprehensive picture of the three-dimensional orientation of the SNR magnetic field can be constructed.
Figure \ref{pa+90} shows the components of the magnetic field in the plane of the sky and along the line of sight as computed from linear- and circular-polarization observations of the OH (1720~MHz) SNR masers.
The line-of-sight component of the magnetic field detected using Zeeman splitting does not change sign throughout the maser regions, consistent with previous observations (cf.\ BFGT).

\subsubsection{Comparison with Observations of Shocked Gas}

Figure \ref{arikawa_linpol} shows the plane-of-the-sky component of the SNR magnetic field indicated by the masers superimposed on an image of the post-shock gas (CO $J = 3 \rightarrow 2$, Arikawa et al.\ 1999).
The magnetic field in the shocked gas is suggested to have been swept and compressed into a plane parallel with the shock (Frail \& Mitchell 1998; Arikawa et al.\ 1999; Dubner et al.\ 2000; Balsara et al.\ 2001).
Figure \ref{arikawa_linpol} indicates that the orientation of the magnetic field in the masers is consistent with the suggested orientation of the magnetic field in the shock.

Considering the good agreement between the orientation of the shocked gas and the orientation of the linear-polarization angle of the maser emission, we feel that our assumption of $RM = 0 \pm 15\ {\rm rad}\,{\rm m}^{-2}$ (\S\ref{RM}) in the direction of the masers is the most physically plausible scenario.

\subsection{Linear Polarization Profiles\label{lin_prof_disc}}

Both the linear-polarization fraction of the maser radiation ($q$) and the position angle of the linearly polarization emission ($\chi$) are observed to vary smoothly as a function of velocity across the maser line profiles; $q$ varies by approximately 1\% and $\chi$ varies by approximately 5\% (Figures \ref{dxdv} and \ref{2pol}; Tables \ref{merpol} and \ref{vlbapol}).
No variation of $q$ or $\chi$ within an OH (1720~MHz) SNR maser has been observed previously, although variations of such small magnitude are not inconsistent with previous analyses.
However, this linear-polarization structure is not expected from the Zeeman splitting or the linear-polarization theory discussed in the previous sections.

A variation in the observed $q$ and $\chi$ values is not necessarily an intrinsic property of the maser emission.
There are two possible explanations for the observed variations:
(1) small changes, as a function of frequency, in the rotation measure and depolarization due to the interstellar medium between the observer and the maser and
(2) Faraday rotation and depolarization {\em within} the gain length of the maser.
In order to quantify the processes in explanation (1), measurements similar to the observations of the synchrotron emission described in \S\ref{lin_disc} would be required.
However, continuum observations would need to have at least 100-mas angular resolution in order to sample adequately the change in sign of $\partial{I}/\partial{\nu}$ between the F39A and F39C masers.
Since applying explanation (1) requires additional observations, we will not discuss it further.
The possible contribution of explanation (2) in OH (1720~MHz) SNR masers has been examined previously.
BFGT find that the gain length of the masers is comparable to the Faraday depolarization length scale, indicating that Faraday rotation cannot contribute significantly to the observed polarization angle.
Similarly, H03 discuss that the observed electron column density (in IC~443) is not sufficient for Faraday rotation to affect significantly the polarization state of the masers.
However, although the maser-gain lengths do not appear to be significantly affected by Faraday processes, we will examine to possibility that explanation (2) may contribute at the few-percent level.

The polarization angle ($\chi$) is determined by the orientation of the plane-of-the-sky component of the magnetic field in the gas (\S \ref{lin_disc}).
In order for the polarization angle to change across the line profile, either of two processes may occur:
(1) a change in the magnetic field orientation in the maser gas as a function of velocity depth into the maser, or
(2) a change in the relationship between the magnetic field in the maser gas and the emergent polarization state.
Since theoretical investigations suggest only a parallel or perpendicular relationship (without a smoothly-varying transition through intermediate angles) between the maser magnetic field and the polarization state ({\it e.g}.\ WW01), explanation (2) is less likely.
In (1), the change in the magnetic field orientation may be in the plane of the sky ($\chi$) or along the line of sight ($\theta$).
However, a large variation of $\theta$ by many tens of degrees would be apparent in our previous Zeeman fitting to $\partial{I}/\partial{\nu}$ (\S 3.3.2).
Using existing models, we can only conclude that
(1) there is a smooth change in $\theta$ or $\chi$ (or both) by about 10 degrees, but that
(2) $\theta$ does not change so much that the relationship between the maser magnetic field and the polarization state transitions from the parallel regime to the perpendicular regime (cf.\ Deguchi \& Watson 1986).

\section{Conclusions}

Using MERLIN, we have imaged all but two of the OH (1720~MHz) SNR masers in W28 with intensity greater than 1~\jb\ as observed using the VLA by Claussen et al.\ (1997).
Of the 28 MERLIN detections, we have detected 4 masers (approximately 15\%) using the VLBA+Y1, consistent with other VLBI detection rates of OH (1720~MHz) SNR masers.
Based on MERLIN data with a resolution of 200~mas and VLBA data with a resolution of 15~mas, the masers have deconvolved sizes of 90 to 350~mas (225 to 875~AU) with compact cores 20~mas (50~AU), consistent with theoretical expectation and previous observations.

Based on a number of constraints, the masers appear to be partially saturated, allowing quantitative theoretical examination of the polarization state of the emission.
These data contain the first direct detection of resolved Zeeman splitting in OH (1720~MHz) SNR masers.
Analysis of the circular-polarization profiles indicates magnetic fields in the maser gas $|\mathbf{B}| \approx 0.75$~mG.
The linear polarization state of the masers is in good agreement with (1) theoretical expectations of the shock geometry, (2) theoretical suggestions concerning the radiative transfer in the masers, and (3) the magnetic field determined using observations of the synchrotron emission from W28, indicating a projected position angle of $-10$\arcdeg\ for the magnetic field in the limb of the SNR near the masers.
Also, the position angle of the magnetic field measured using the linear polarization of the masers is in good agreement with both the orientation of the shock observed using thermal molecular emission ({\it e.g}.\ Frail \& Mitchell 1998) and with the magnetic field orientation expected from the SNR expansion ({\it e.g}.\ Balsara et al.\ 2001).

The polarization properties of the masers are observed to change as a function of velocity within the profile, indicating small (approximately 10\arcdeg) changes in the magnetic field orientation within the maser gas.
Using theoretical models, the angle between the line-of-sight and the magnetic field vector is inferred to be $\theta \approx 60\arcdeg$.

\acknowledgments

\begin{figure}
\hbox{ \epsfxsize=4.50in \epsfbox{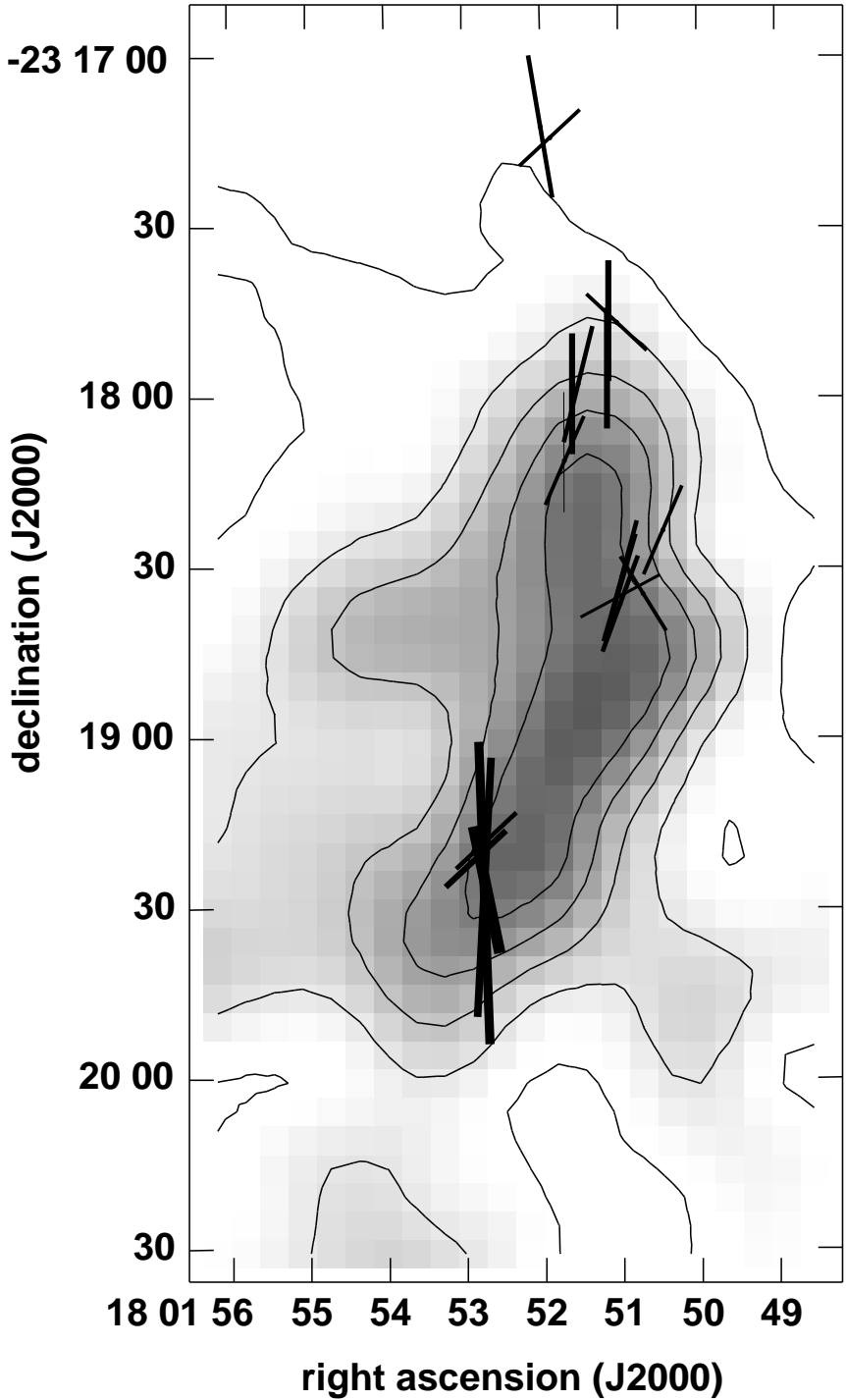}}
\caption{Plot of the position angles of the magnetic field from Figure \ref{pa+90} overlayed on a greyscale and contour moment-zero image of the shocked CO ($J = 3 \rightarrow 2$) emission from Arikawa et al. (1999; see Fig. 1).
The contour levels are 8, 16, 24, 32, and 40 times $2.67 \times 10^4\ {\rm K}\,{\rm m}\,{\rm s}^{-1}$.
The angular resolution of the CO image is 15\arcsec\ and the absolute positional accuracy of the CO image is about 1\arcsec.
\label{arikawa_linpol}}
\end{figure}

IMH was supported by the NRAO pre-doctoral researcher program.
We thank M.\ Elitzur and W.\ D.\ Watson for a number of helpful discussion and insights regarding the manuscript.
We thank Y.\ Arikawa and K.\ Tatematsu for communicating the CO data used in Figures 1 and \ref{arikawa_linpol}.
We thank A.\ M.\ S.\ Richards for assistance with reduction of the MERLIN observations and P.\ Thomasson for flexible scheduling of the MERLIN array.
We thank the anonymous referee for providing comments which significantly improved the clarity of the presentation and the interpretation of the maser theories.

\clearpage

\clearpage

\clearpage

\clearpage

\clearpage

\clearpage

\clearpage

\clearpage

\clearpage

\clearpage

\clearpage

\clearpage

\clearpage

\clearpage

\clearpage

\clearpage

\begin{figure}
\plotone{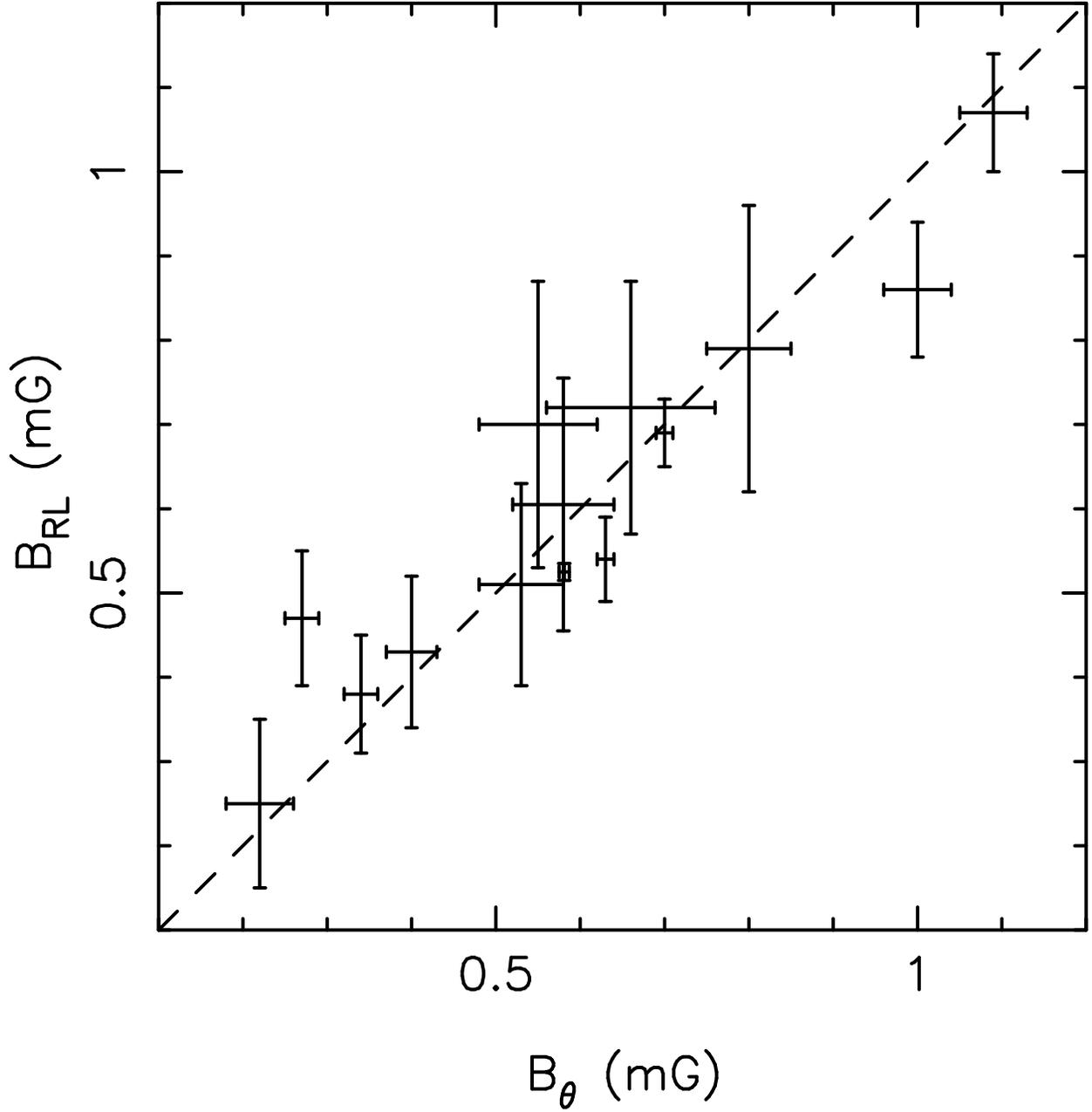}
\caption{Plot of $B_{\rm RL}$ versus $B_\theta$ for the magnetic field strengths fitted to the MERLIN data (Table \ref{circ}).
Each point is a separate maser.
The dashed line has a slope of one indicating the good agreement between $B_{\rm RL}$ and $B_\theta$.
\label{MERLIN_RLD}}
\end{figure}

\clearpage

\begin{deluxetable}{ r@{\ }r@{\ }l  r@{\ }r@{\ }l }
\tablecaption{VLBA Observation Positions}
\tablecolumns{6}
\tablewidth{0pt}
\tablehead{
\multicolumn{3}{c}{R.A.\ (J2000)} & \multicolumn{3}{c}{Decl.\ (J2000)} \\
\colhead{(h} & \colhead{m} & \colhead{s)} & \colhead{(\arcdeg} & \colhead{\arcmin} & \colhead{\arcsec)}
}
\startdata
\cutinhead{array tracking positions}
18&00&57.5393 & -23&17&02.459 \\
  &01&52.7056 &    &19&24.639 \\
\cutinhead{correlation centers}
  &00&43.7492 &    &17&28.865 \\
  &01&15.2832 &    &01&04.384 \\
  &01&37.9266 &    &28&51.416 \\
  &01&44.3510 &    &26&21.445 \\
  &01&50.8221 &    &18&25.577 \\
  &01&51.4640 &    &18&04.530 \\
  &01&51.5999 &    &17&27.719 \\
  &01&52.7056 &    &19&24.639 
\enddata
\end{deluxetable}

\clearpage

\begin{deluxetable}{ l  r@{.}l  r@{.}l@{$\times$}r@{.}l c }
\tablecaption{MERLIN Maser Sizes and Flux Densities\label{mersizes}}
\tablecolumns{8}
\tablewidth{0pt}
\tablehead{
\colhead{Feature} & \multicolumn{2}{c}{$S$ (Jy)} & \multicolumn{4}{c}{angular size\tablenotemark{a}\ (mas)} & \colhead{P.A.\ (deg)}
}
\startdata
B7   &  0&25 & \multicolumn{4}{c}{\tablenotemark{b}} & \tablenotemark{b} \\
D13  &  0&82 & \multicolumn{4}{c}{\tablenotemark{b}} & \tablenotemark{b} \\
F14  &  2&13 & 0&26(1)   &  0&52(3)  & 74(8)   \\
F16  &  0&51 & $<0$&19   &  $<0$&17  & \tablenotemark{b} \\
F19  &  0&78 & \multicolumn{4}{c}{\tablenotemark{b}} & \tablenotemark{b} \\
F20A &  1&94 & 0&35(4)   &  0&75(4)  & 41(8)   \\
F20B &  1&66 & 0&32(3)   &  0&54(4)  & 43(7)   \\
F21  &  3&58 & 0&3(1)    &  0&62(3)  & 153(15) \\
E22  &  2&76 & 0&45(1)   &  0&51(1)  & 17(3)   \\
E23  &  0&66 & 0&20(6)   &  0&37(8)  & 42(16)  \\
E24  &  6&90 & 0&10(1)   &  $<0$&45  & \tablenotemark{b} \\
E26  &  0&21 & 0&4(3)    &  $<0$&3   & \tablenotemark{b} \\
E27  &  0&40 & 0&21(1)   &  0&49(9)  & 55(30)  \\
E30  &  3&83 & 0&31(3)   &  0&37(4)  & 133(5)  \\
E31  & 12&22 & 0&26(1)   &  0&43(1)  & 96(4)   \\
E33  &  2&80 & 0&25(2)   &  0&56(5)  & 92(15)  \\
E34  &  0&95 & 0&25(5)   &  0&52(6)  & 45(15)  \\
E35  &  0&51 & 0&60(5)   &  0&53(9)  & 21(6)   \\
E36  &  3&13 & 0&31(1)   &  0&66(5)  & 72(12)  \\
F39A & 45&67 & 0&092(1)  &  0&290(3) & 100(4)  \\
F39B &  3&36 & 1&0(1)    &  0&2(1)   & 3(4)    \\
F39C & 23&25 & 0&097(2)  &  0&14(1)  & 51(3)   \\
F39D &  4&10 & 0&45(1)   &  0&27(2)  & 21(1)   \\
F40  &  0&88 & 0&14(6)   &  0&3(2)   & 73(30)  \\
F41  &  1&46 & 0&30(5)   &  0&32(5)  & 16(7)   
\enddata
\tablenotetext{a}{the deconvolved angular size of the masers listed (major axis)$\times$(minor axis)}
\tablenotetext{b}{insufficient signal-to-noise for deconvolution}
\end{deluxetable}

\clearpage

\begin{deluxetable}{ l  r@{.}l r r}
\tablecaption{VLBA Maser Sizes and Flux Densities\label{vlbasizes}}
\tablecolumns{5}
\tablewidth{0pt}
\tablehead{
\colhead{Feature} & \multicolumn{2}{c}{$S$ (Jy)} & \colhead{angular size (mas)} & \colhead{P.A.\ (deg)}
}
\startdata
E24     &   1&50 & $<4$ & \tablenotemark{a} \\
F39AI   &  27&10 & 14(2)& 4(2) \\ 
F39AII  &  10&30 & 14(4)& 4(3) \\ 
F39AIII &   2&70 &    \tablenotemark{a} & \tablenotemark{a} \\ 
F39CI   &   8&90 & $<4$ & \tablenotemark{a} \\
F39CII  &   7&70 & $<4$ & \tablenotemark{a} \\
F40     &   0&50 &    \tablenotemark{a} & \tablenotemark{a} 
\enddata 
\tablenotetext{a}{insufficient signal-to-noise for deconvolution}
\end{deluxetable}

\clearpage

\begin{deluxetable}{ l  r@{\ }r@{\ }l  r@{\ }r@{\ }l  c  r@{.}l  r@{.}l  r@{.}l }
\tablecaption{MERLIN Fitted Maser Properties\label{merpos}}
\tablecolumns{14}
\tablewidth{0pt}
\tablehead{
\colhead{Feature} & \multicolumn{3}{c}{R.A.\ (J2000)} & \multicolumn{3}{c}{Decl.\ (J2000)} & \colhead{$I$} & \multicolumn{2}{c}{$\Delta{v}$} & \multicolumn{2}{c}{$v_{\rm LSR}$} & \multicolumn{2}{c}{$T_B$} \\
 & \colhead{(h} & \colhead{m} & \colhead{s)} & \colhead{(\arcdeg} & \colhead{\arcmin} & \colhead{\arcsec)} & \colhead{(\mjb)} & \multicolumn{2}{c}{(\kms)} & \multicolumn{2}{c}{(\kms)} & \multicolumn{2}{c}{($10^6$~K)} \\
}
\startdata
B7   &   &01&15.637  &      &16&36.67  & 150(50)\tablenotemark{a} & 1&34(8) &  5&22(3) &   1&3  \\
D13  &   &  &44.349  &      &26&21.24  & 270(50)\tablenotemark{a} & 0&89(3) & 13&96(1) &   2&3  \\
F14  &   &  &50.754  &      &18&20.76  &  1120      & 1&04(1) & 15&36(1)  &   9&5  \\
F16  &   &  &50.977  &      &  &32.07  &   469      & 0&85(2) & 11&95(1)  &   4&0  \\
F19  &   &  &51.210  &      &  &29.09  &   287      & 0&67(2) & 11&21(1)  &   2&4  \\
F20A &   &  &51.227  &      &  &33.85  &   712      & 1&11(2) &  9&27(1)  &   6&1  \\
F20B &   &  &51.229  &      &  &32.47  &   721      & 0&96(2) &  8&92(1)  &   6&1  \\
     &   &  &        &      &  &       &   232      & 1&00(4) & 10&28(2) &   2&0  \\
F21  &   &  &51.238  &      &  &30.98  &   895      & 0&89(2) & 10&20(1)  &   7&6  \\
E22  &   &  &51.269  &      &17&43.89  &   863      & 1&00(2) & 15&29(1)  &   7&6  \\
E23  &   &  &51.341  &      &  &45.61  &   479      & 0&96(3) & 15&69(2) &   4&1  \\
E24  &   &  &51.350  &      &  &43.50  &  3214      & 0&94(1)  & 13&10(1)  &  27&4  \\
E26  &   &  &51.376  &      &  &51.95  &   167      & 0&91(5) & 11&21(2) &   1&4  \\
E27  &   &  &51.415  &      &  &47.56  &   191      & 1&22(4) & 12&70(2) &   1&6  \\
     &   &  &        &      &  &       &    97      & 0&77(6) & 11&22(3) &   0&8  \\
E30  &   &  &51.697  &      &  &54.89  &  1571      & 0&78(1)  & 11&62(1)  &  13&4  \\
E31  &   &  &51.755  &      &  &56.60  &  3003      & 0&95(1)  & 11&68(1)  &  25&6  \\
E33  &   &  &51.850  &      &18&08.46  &   585      & 0&94(2) &  9&37(1)  &   5&0  \\
E34  &   &  &51.852  &      &  &07.00  &   398      & 1&11(2) & 11&23(1) &   3&4  \\
E35  &   &  &52.015  &      &17&11.16  &   232      & 0&87(3) & 11&92(1)  &   2&0  \\
E36  &   &  &52.114  &      &  &09.17  &   869      & 0&75(1)  & 11&56(1)  &   7&4  \\
F39A &   &  &52.707  &      &19&24.65  & 28890      & 0&70(1)  & 11&18(1)  & 246&0  \\
F39B &   &  &52.707  &      &  &25.55  &  1221      & 0&73(2) &  9&78(1)  &  10&4  \\
F39C &   &  &52.731  &      &  &24.20  & 19279      & 0&77(1)  &  9&75(1)  & 164&2  \\
F39D &   &  &52.736  &      &  &25.21  &  2579      & 0&84(2) & 10&79(1)  &  22&0  \\
F40  &   &  &52.714  &      &  &15.94  &   604      & 0&69(2) &  8&96(1)  &   5&1  \\
F41  &   &  &52.826  &      &  &19.18  &  1265      & 0&76(1)  &  9&94(1)  &  10&8  
\enddata
\tablenotetext{a}{corrected for beam attenuation and time smearing}
\end{deluxetable}

\clearpage

\begin{deluxetable}{ l  r@{\ }r@{\ }l  r@{\ }r@{\ }l  c  r@{.}l  r@{.}l  r@{.}l }
\tablecaption{VLBA Fitted Maser Properties\label{vlbapos}}
\tablecolumns{14}
\tablewidth{0pt}
\tablehead{
\colhead{Feature} & \multicolumn{3}{c}{R.A.\ (J2000)} & \multicolumn{3}{c}{Decl.\ (J2000)} & \colhead{$I$} & \multicolumn{2}{c}{$\Delta{v}$} & \multicolumn{2}{c}{$v_{\rm LSR}$} & \multicolumn{2}{c}{$T_B$} \\
 & \colhead{(h} & \colhead{m} & \colhead{s)} & \colhead{(\arcdeg} & \colhead{\arcmin} & \colhead{\arcsec)} & \colhead{(\mjb)} & \multicolumn{2}{c}{(\kms)} & \multicolumn{2}{c}{(\kms)} & \multicolumn{2}{c}{($10^9$~K)} \\
}
\startdata
E24     & 18&01&51.2952 & $-23$&17&43.480 &   410      & 0&73(4)   & 13&26(2)   &   0&8  \\
F39AIII &   &  &52.7026 &      &19&24.723 &  1650      & 0&70(1)   & 11&27(1)  &   3&5  \\
F39AII  &   &  &52.7038 &      &  &24.655 &  1790      & 0&57(1)  & 11&23(1)  &   5&8  \\
F39AI   &   &  &52.7054 &      &  &24.641 &  2440      & 0&50(1)  & 11&20(1)  &   8&9  \\
F39CII  &   &  &52.7267 &      &  &24.239 &  1630      & 0&55(2) &  9&67(1)  &   3&2  \\
F39CI   &   &  &52.7287 &      &  &24.212 &  1300      & 0&65(1)  &  9&74(1)  &   2&5  
\enddata
\end{deluxetable}

\clearpage

\begin{deluxetable}{ l  r@{.}l r@{.}l@{$\pm$}r@{.}l r@{.}l@{$\pm$}r@{.}l   r@{.}l r@{.}l@{$\pm$}r@{.}l r@{.}l@{$\pm$}r@{.}l }
\tablecaption{Zeeman Results\label{circ}}
\tablecolumns{13}
\tablewidth{0pt}
\tablehead{
\colhead{Feature} & \multicolumn{10}{c}{MERLIN} & \multicolumn{10}{c}{VLBA} \\ \cline{3-10} \cline{13-20}
      & \multicolumn{2}{c}{$v_{\rm RCP} - v_{\rm LCP}$\tablenotemark{a}} & \multicolumn{4}{c}{$B_{\rm RL}$} & \multicolumn{4}{c}{$B_\theta$} & \multicolumn{2}{c}{$v_{\rm RCP} - v_{\rm LCP}$\tablenotemark{a}} & \multicolumn{4}{c}{$B_{\rm RL}$} & \multicolumn{4}{c}{$B_\theta$} \\
      & \multicolumn{2}{c}{(km/s)} & \multicolumn{4}{c}{(mG)} & \multicolumn{4}{c}{(mG)} & \multicolumn{2}{c}{(km/s)} & \multicolumn{4}{c}{(mG)} & \multicolumn{4}{c}{(mG)}
}
\startdata
F14     &   0&029(12) & 0&25 & 0&10 & 0&22&0&04  & \multicolumn{2}{c}{\tablenotemark{b}} & \multicolumn{4}{c}{\tablenotemark{b}} & \multicolumn{4}{c}{\tablenotemark{b}} \\
F16     &   0&082(17) & 0&72 & 0&15 & 0&66&0&10  & \multicolumn{2}{c}{\tablenotemark{b}} & \multicolumn{4}{c}{\tablenotemark{b}} & \multicolumn{4}{c}{\tablenotemark{b}} \\
F19     & \multicolumn{6}{c}{\ldots}& 0&34&0&14  & \multicolumn{2}{c}{\tablenotemark{b}} & \multicolumn{4}{c}{\tablenotemark{b}} & \multicolumn{4}{c}{\tablenotemark{b}} \\
F20A    &   0&08(2)   & 0&70 & 0&17 & 0&55&0&07  & \multicolumn{2}{c}{\tablenotemark{b}} & \multicolumn{4}{c}{\tablenotemark{b}} & \multicolumn{4}{c}{\tablenotemark{b}} \\
F20B    &   0&070(17) & 0&61 & 0&15 & 0&58&0&06  & \multicolumn{2}{c}{\tablenotemark{b}} & \multicolumn{4}{c}{\tablenotemark{b}} & \multicolumn{4}{c}{\tablenotemark{b}} \\
F21     &   0&058(14) & 0&51 & 0&12 & 0&53&0&05  & \multicolumn{2}{c}{\tablenotemark{b}} & \multicolumn{4}{c}{\tablenotemark{b}} & \multicolumn{4}{c}{\tablenotemark{b}} \\
E22     &   0&09(2)   & 0&79 & 0&17 & 0&80&0&05  & \multicolumn{2}{c}{\tablenotemark{b}} & \multicolumn{4}{c}{\tablenotemark{b}} & \multicolumn{4}{c}{\tablenotemark{b}} \\
E23     & \multicolumn{6}{c}{\ldots}& 0&27&0&09  & \multicolumn{2}{c}{\tablenotemark{b}} & \multicolumn{4}{c}{\tablenotemark{b}} & \multicolumn{4}{c}{\tablenotemark{b}} \\
E24     &   0&062(6)  & 0&54 & 0&05 & 0&63&0&01  &                                                       0&051(34) & 0&45 & 0&27 & 0&36&0&14 \\
E26     & \multicolumn{6}{c}{\ldots}& 1&15&0&20  & \multicolumn{2}{c}{\tablenotemark{b}} & \multicolumn{4}{c}{\tablenotemark{b}} & \multicolumn{4}{c}{\tablenotemark{b}} \\
E27     & \multicolumn{6}{c}{\ldots}& 0&84&0&16  & \multicolumn{2}{c}{\tablenotemark{b}} & \multicolumn{4}{c}{\tablenotemark{b}} & \multicolumn{4}{c}{\tablenotemark{b}} \\
E30     &   0&049(10) & 0&43 & 0&09 & 0&40&0&03  & \multicolumn{2}{c}{\tablenotemark{b}} & \multicolumn{4}{c}{\tablenotemark{b}} & \multicolumn{4}{c}{\tablenotemark{b}} \\
E31     &   0&043(8)  & 0&38 & 0&07 & 0&34&0&02  & \multicolumn{2}{c}{\tablenotemark{b}} & \multicolumn{4}{c}{\tablenotemark{b}} & \multicolumn{4}{c}{\tablenotemark{b}} \\
E33     & \multicolumn{6}{c}{\ldots}& 0&53&0&08  & \multicolumn{2}{c}{\tablenotemark{b}} & \multicolumn{4}{c}{\tablenotemark{b}} & \multicolumn{4}{c}{\tablenotemark{b}} \\
E35     & \multicolumn{6}{c}{\ldots}& 0&65&0&19  & \multicolumn{2}{c}{\tablenotemark{b}} & \multicolumn{4}{c}{\tablenotemark{b}} & \multicolumn{4}{c}{\tablenotemark{b}} \\
F39AI   &   0&060(1)  & 0&52 & 0&01 &0&581&0&006 &                                                       0&042(4)  & 0&37 & 0&03 & 0&31&0&02 \\
F39AII  & \multicolumn{10}{c}{\ldots}            &                                                       0&059(5)  & 0&52 & 0&04 & 0&47&0&04 \\
F39AIII & \multicolumn{10}{c}{\ldots}            &                                                       0&064(9)  & 0&56 & 0&08 & 0&37&0&09 \\
F39B    &   0&098(18) & 0&86 & 0&08 & 1&00&0&04  & \multicolumn{2}{c}{\tablenotemark{b}} & \multicolumn{4}{c}{\tablenotemark{b}} & \multicolumn{4}{c}{\tablenotemark{b}} \\
F39CI   &   0&079(5)  & 0&69 & 0&04 & 0&70&0&01  &                                                       0&093(8)  & 0&82 & 0&07 & 0&63&0&08 \\
F39CII  & \multicolumn{10}{c}{\ldots}            &                                                       0&075(9)  & 0&66 & 0&08 & 0&50&0&06 \\
F39D    &   0&054(9)  & 0&47 & 0&08 & 0&27&0&02  & \multicolumn{2}{c}{\tablenotemark{b}} & \multicolumn{4}{c}{\tablenotemark{b}} & \multicolumn{4}{c}{\tablenotemark{b}} \\
F41     &   0&122(8)  & 1&07 & 0&07 & 1&09&0&04  & \multicolumn{2}{c}{\tablenotemark{b}} & \multicolumn{4}{c}{\tablenotemark{b}} & \multicolumn{4}{c}{\tablenotemark{b}} 
\enddata
\tablenotetext{a}{the velocity difference between the centers of the Gaussians fitted to the RCP and LCP line profiles}
\tablenotetext{b}{feature not detected}
\end{deluxetable}

\clearpage

\begin{deluxetable}{ l  c  c  c  r  c }
\tablecaption{MERLIN Linear Polarization Results\label{merpol}}
\tablecolumns{6}
\tablewidth{0pt}
\tablehead{
\colhead{Feature} & \colhead{$Q$} & \colhead{$U$} & \colhead{$\chi$} & \colhead{$q$} & \colhead{$\partial{\chi}/\partial{\nu}$} \\
 & \colhead{(\mjb)} & \colhead{(\mjb)} & \colhead{(deg)} & \colhead{(\%)} & \colhead{(deg/Hz)} \\
}
\startdata
B7   &   $<10$ &  $<10$ &    -        & $<70$ &\tablenotemark{a} \\
D13  &   $<10$ &  $<10$ &    -        & $<40$ &\tablenotemark{a} \\
F14  &   $-17$ &  $ 17$ & $ 68$(12)   &  2(1) &\tablenotemark{a} \\
F16  &   $-15$ &  $-25$ & $-60$(12)   &  6(2) &\tablenotemark{a} \\
F19  &   $-29$ &  $ 15$ & $ 76$(10)   & 11(4) &\tablenotemark{a} \\
F20A &   $-33$ &  $ 25$ & $ 71$(8)    &  6(1) &   $-0.0023$(10)  \\
F20B &   $ 17$ &  $ 30$ & $ 30$(8)    &  5(1) &   $-0.0043$(20)  \\
     &   $<10$ &  $ 17$ & $ 45$(9)    &  9(2) &   $-0.0022$(10)  \\
F21  &   $-41$ &  $ 24$ & $ 75$(6)    &  5(1) &      $0.012$(8)  \\
E22  &   $<10$ &  $-18$ & $-45$(9)    &  3(1) &\tablenotemark{a} \\
E23  &   $<10$ &  $<10$ &    -        & $<6$  &\tablenotemark{a} \\
E24  &   $-52$ &  $<10$ & $90 $(3)    &  2(1) &\tablenotemark{a} \\
E26  &   $-35$ &  $<10$ & $90 $(5)    & 22(5) &\tablenotemark{a} \\
E27  &   $<10$ &  $<10$ &    -        & $<10$ &\tablenotemark{a} \\
     &   $<10$ &  $<10$ &    -        & $<20$ &\tablenotemark{a} \\
E30  &   $-30$ &  $-15$ & $ 77$(8)    &  2(1) &\tablenotemark{a} \\
E31  &   $-48$ &  $<10$ & $90 $(4)    &  2(1) &\tablenotemark{a} \\
E33  &   $-12$ &  $ 12$ & $ 68$(16)   &  2(1) &\tablenotemark{a} \\
E34  &   $ 11$ &  $<10$ & $90 $(11)   &  4(1) &\tablenotemark{a} \\
E35  &   $<10$ &  $ 25$ & $ 45$(7)    & 12(2) &\tablenotemark{a} \\
E36  &   $-25$ &  $ -8$ & $-81$(9)    &  3(1) &\tablenotemark{a} \\
F39A & $-1712$ & $-689$ & $-79.0$(13) &  7(1) &   $-0.00236$(6)  \\
F39B &  $-109$ &  $-37$ & $-81$(3)    &  9(1) &\tablenotemark{a} \\
F39C & $-2159$ & $ 214$ & $ 87.2$(35) & 11(1) &    $0.00076$(9)  \\
F39D & $ -139$ & $ -11$ & $-88$(2)    &  5(1) &   $-0.0053$(11)  \\
F40  &   $<10$ & $  33$ & $ 45$(5)    &  6(1) &\tablenotemark{a} \\
F41  &   $<10$ & $  77$ & $ 45$(3)    &  6(1) &    $0.0040$(15)  
\enddata
\tablenotetext{a}{insufficient signal-to-noise to constrain slope}
\tablenotetext{b}{no Zeeman profile observed}
\end{deluxetable}

\clearpage

\begin{deluxetable}{ l  c  c  c  r  c }
\tablecaption{VLBA Linear Polarization Results\label{vlbapol}}
\tablecolumns{6}
\tablewidth{0pt}
\tablehead{
\colhead{Feature} & \colhead{$Q$} & \colhead{$U$} & \colhead{$\chi$} & \colhead{$q$} & \colhead{$\partial{\chi}/\partial{\nu}$} \\
 & \colhead{(\mjb)} & \colhead{(\mjb)} & \colhead{(deg)} & \colhead{(\%)} & \colhead{(deg/Hz)} \\
}
\startdata
E24     & $<30$  & $<30$  & -        & $<15$ &\tablenotemark{a} \\
F39AI   & $-190$ & $ -70$ & $-81$(6) &  8(1) &    $-0.0038$(14) \\
F39AII  & $-170$ &  $<30$ & $90 $(8) &  9(1) &\tablenotemark{a} \\
F39AIII &  $<30$ &  $<30$ &  -       & $<6$  &\tablenotemark{a} \\
F39CI   & $-300$ & $  90$ & $ 82$(7) & 24(2) &     $0.0058$(16) \\
F39CII  & $-280$ & $ 220$ & $ 71$(8) & 22(2) &\tablenotemark{a} 
\enddata
\tablenotetext{a}{insufficient signal-to-noise to constrain slope}
\end{deluxetable}

\end{document}